\documentclass[preprint,12pt,3p]{elsarticle}
\usepackage{hyperref}
\usepackage{graphicx}
\usepackage{epstopdf}
\usepackage[utf8]{inputenc}
\usepackage{amsmath}
\usepackage{bm}
\usepackage{mathtools}
\usepackage{setspace}
\font\grassettogreco=cmmib10
\font\scriptgrassettogreco=cmmib7
\font\scriptscriptgrassettogreco=cmmib10 at 5 truept
\font\sansserif=cmss10
\font\scriptsansserif=cmss10 at 7 truept
\font\scriptscriptsansserif=cmss10 at 5 truept
\def\bgr{\fam=13}
\def\ssm{\fam=14}
\input amssym.def
\input amssym.tex
\def \ebf{{\bf e}}

\def \xbf{{\bf x}}
\def \ybf{{\bf y}}
\def \wbf{{\bf w}}

\def \Mop{{\mathchardef\alpha="710B \ssm \char'115}}

\def \Pop{{\mathchardef\alpha="710B \ssm \char'120}}

\def \Rop{{\mathchardef\alpha="710B \ssm \char'122}}

\def\Xibf{{\mathchardef\Xi="7104 \bgr \Xi}}

\def\xibf{{\mathchardef\xi="7118 \bgr \xi}}

\def \Acal{{\cal A}}
\def \Bcal{{\cal B}}

\def \Ecal{{\cal E}}

\def \Hcal{{\cal H}}

\def \Lcal{{\cal L}}
\def \Mcal{{\cal M}}

\def \eps{\epsilon}

  2
  3
 4
 5

\def \spazio{\vskip .5	  truecm \noindent }

\def \spa{\vskip    .3	  truecm \noindent }

\def\spi{\vskip .15	 truecm  \noindent}
\def \pan {\par \noindent}

   \newcount\Silicon \global\Silicon=1
   \newcount\PC \global\PC=2
   \newcount\parziale
   \def\Colori#1{\global\parziale=#1
                \ifnum\parziale=\Silicon
                    \input colors
                    \gdef\Color##1{\Black{##1}}                    
                \else\ifnum\parziale=\PC
                    \input colordvi
                    \gdef\textRGB##1{\textColor{##1 0.}}
                    \gdef\GrayA##1{\textGray{##1}}
                    \gdef\GrayB##1{\textGray{##1}}
                    \gdef\GrayC##1{\textGray{##1}}
                    \gdef\GrayD##1{\textGray{##1}}
                    \gdef\GrayE##1{\textGray{##1}}
                    \gdef\GrayF##1{\textGray{##1}}
                    \gdef\GrayG##1{\textGray{##1}}
                    \gdef\GrayH##1{\textGray{##1}}\fi                   
                \fi}        
\def \parton#1{\left({{#1}}\right)}
\def \parqua#1{\left[{{#1}}\right]}

\font \rmsmm=cmr7
\def \hbix#1{\rmsmm \hbox{ {#1} } }

\def \underwrite#1{\mathop{\vtop{\ialign {##\crcr
$\hfil\displaystyle {#1}\hfil$\crcr\noalign{\kern3pt\nointerlineskip}
\crcr\noalign{\kern3pt}}}} \limits}

\def\Reali{\Bbb R}

\def\Toro{\Bbb T}

\def\sqr#1#2{{\vcenter{\hrule height.#2pt
     \hbox{\vrule width.#2pt height#1pt \hskip#1pt
       \vrule width.#2pt}
     \hrule height.#2pt}}}

\def\DAL{\hbox{\raise.250ex \hbox{$\sqr7{10}\,$}}} 

\def\bib#1.#2/#3/#4/#5/#6/#7.{\frenchspacing\item{[#1]}#2:\ {\it ``#3''}
~--~#4\ $\underline{\bf #5}$,\ #6 (#7)}
\def\diagramma#1#2#3#4#5#6#7#8{
 \vbox to 2.5cm{
       \hbox to 3cm{\hfil ${#1}$ \hfil}
       \hbox to 3cm{ ${#2}$\rightarrowfill ${#3}$ }
\hbox to 3cm{${#4} \Biggl\uparrow\hfill\Biggr\uparrow {#5}$}
       \hbox to 3cm{ ${#6}$\rightarrowfill ${#7}$ }
       \hbox to 3cm{\hfil ${#8}$ \hfil}   }    }
\def\sopra#1#2{{\raise 0.8 ex
\hbox{$
{{\scriptstyle \,{#2}}	\atop \displaystyle{#1}}
$}}
}

\def\figuraps#1#2#3#4#5{
\par
\midinsert
\centerline{\bf #4}
\vbox to #3 truecm{
\vskip #3 truecm
\ifnum #1 = 0	
\special {ps: plotfile #2}
\else		
\special {#2 0 0 moveto 16} \fi
}
\centerline{#5}
\endinsert}

\font\cofon = cmr6
\font\cobfon = cmbx6
\font\copi = cmr9
\def\codlib{{\copi\copyright}{\cofon 88-08- }{\cobfon 9820}}
\def\riga{\vskip .1  truecm   \hrule \vskip .2	    truecm \noindent }
\def\HeadLinea#1#2{	\headline={\vbox to 0pt{\vss\noindent
{\ifnum \pageno=1  \hfill {\bf \folio}		 
\else {\ifodd \pageno			   
{\noindent     \hfill  {\it     #2} \quad {\bf \folio}
 }\riga
\else				 
{\noindent {\bf\folio} \quad  {\it  #1} \hfill 	}
\riga
\fi } \fi }			}}
}
\def\testatina#1#2{	\headline={\vbox to 0pt{\vss\noindent
{\ifnum \pageno=1  \hfill {\bf \folio}		 
\else {\ifodd \pageno			   
{\noindent  \codlib   \hfill  {\it     #2} \quad {\bf \folio}
 }\riga
\else				 
{\noindent {\bf\folio} \quad  {\it  #1} \hfill \codlib	}
\riga
\fi } \fi }			}}
}
\def\testatinacap#1#2#3{	\headline={\vbox to 0pt{\vss\noindent
{\ifnum \pageno=#3  \hfill {\bf \folio}		 
\else {\ifodd \pageno			   
{\noindent  \codlib   \hfill  {\it     #2} \quad {\bf \folio}
 }\riga
\else				 
{\noindent {\bf\folio} \quad  {\it  #1} \hfill \codlib	}
\riga
\fi } \fi }			}}
}
\def\oggi{\number\day\space\ifcase\month
   \or gennaio\or febbraio\or marzo\or aprile\or maggio\or giugno\or
   luglio\or agosto\or settembre\or ottobre\or novembre\or dicembre
   \fi\space\number\year}
\def\today{\number\day\space\ifcase\month
   \or January\or February\or March\or April\or May\or June\or
   July\or August\or September\or October\or November \or December
   \fi\space\number\year}   
\def\frame#1{\ifmmode\dframe{#1}\else\leavevmode\lower 2.4 pt
    \hbox{\vrule\unskip\vbox{\hrule\kern 1.5 pt\hbox{\kern
    1.5 pt{#1}\kern 0.5 pt}\kern 2 pt\hrule}\unskip\vrule}\fi}

\def\dframe#1{\hbox{\vrule\unskip$\vcenter{\hrule\kern 3 pt\hbox
    {\kern 3 pt$\displaystyle{#1}$\kern3pt}\kern 3 pt\hrule}$\vrule}}

\def\bm{\bf}
\def\Tr{\hbox{Tr}\,}
\hypersetup{unicode=true,colorlinks=true, urlcolor=red,linkcolor=blue,citecolor=blue,filecolor=blue}

\usepackage{amsthm}

\journal{Commun Nonlinear Sci Numer Simulat}

\begin{document}

\begin{frontmatter}

\title{Fidelity and Reversibility in\\
the Restricted Three Body Problem}

\author[label1]{F. Panichi\corref{cor1}}
\address[label1]{Institute of Physics and CASA*, University of Szczecin,\\
ul. Wielkopolska 15, PL-70-451 Szczecin, Poland}

\cortext[cor1]{I am corresponding author}

\ead{federico.panichi@studio.unibo.it}
\author[label5]{L. Ciotti}
\author[label5]{G. Turchetti}
\address[label5]{Department of Physics and Astronomy, Alma Mater Studiorum - University of Bologna, \\
  Viale Berti Pichat, 6/2 - 40127 Bologna, Italy}

\begin{abstract}
{\small We use the Reversibility Error Method and the Fidelity to analyze
the global effects of a small perturbation in a non-integrable
system. Both methods have already been proposed and used in the
literature but the aim of this paper is to compare them in a
physically significant example adding some considerations on the
equivalence, observed in this case, between round-off and random
perturbations.

As a paradigmatic example we adopt the restricted planar circular three body
problem.  The cumulative effect of random perturbations or round-off leads to
a divergence of the perturbed orbit from the reference one.  Rather
than computing the distance of the perturbed orbit from the reference
one, after a given number $n$ of iterations, a procedure we name the
Forward Error Method (FEM), we measure the distance of the reversed
orbit ($n$ periods forward and backward) from the initial point.
This approach, that we name Reversibility Error Method (REM), does not
require the computation of the unperturbed map.  The loss of memory of
the perturbed map is quantified by the Fidelity decay rate whose computation
requires a statistical average over an invariant region. Two distinct
definitions of Fidelity are given. The asymptotic
equivalence of REM and FEM is analytically proved for linear symplectic
maps with random perturbations.  For a given map, the REM plot provides
a picture of the dynamic stability regions in the phase space, very
easy to obtain for any kind of perturbation and very simple to
implement numerically.  The REM and FEM for linear
symplectic maps are proved to be asymptotically equivalent.
The global error growth follows a power law in the
regions of integrable (or quasi integrable) motion and an exponential
law in the regions of chaotic motion. We prove that the power law exponent
is $3/2$ for a generic anisochronous system, but drops down to $1/2$ if
the system is isochronous.  Correspondingly the Fidelity $F(t)$
exhibits an exponential decay and $-\ln F(t)$ grows just as the square
of the FEM or REM error.  The Reversibility Error and Fidelity
can be used for a quantitative analysis of dynamical systems and are
suited to investigate the transition regions from chaotic to
regular motion even for Hamiltonian systems with many degrees of
freedom such as the $N$-body problem.}

\end{abstract}

\end{frontmatter}
\newpage
\section{Introduction}
\def\spi {\vskip 0.1 truecm }
\label{sec1}
\noindent
The dynamics of non-integrable Hamiltonian systems is qualitatively
well understood when it reduces to an area preserving map on the
Poincaré section. This is the case of the $3$-body problem. However
the orbits, for generic initial conditions, can only be obtained by
numerical integration \cite{AarsethBook,NbodyLecture}. The
symplectic integration schemes preserve the Poincaré invariants
\cite{SymplNbody,SymplNbody2} but are affected by local
discretization errors and round-off (\cite{errors,olderror}
and reference therein for a recent review on the topic). Therefore, it
is important to be able to estimate the divergence of the numerically
integrated orbit with respect to the exact one.  With a sufficiently
small integration step the local integration error can be lowered down
close to the round-off level. For a recent and comprehensive
presentation of geometric integration methods, their accuracy and
stability see \cite{Hairer}.  By slightly changing the initial point
in phase space we explore the sensitivity of the map to initial
conditions.  The Lyapunov Error Method (LEM) is based on the distance
of the reference orbit from another orbit with a close initial
condition. The definition is given at the beginning of Section 3.  The
asymptotic analysis provides the Lyapunov spectrum.  The
rigorous analysis developed by \cite{Oseledec,Benettin} has a
numerical counterpart \cite{Benettin2},  and several methods to explore
the asymptotic behavior of LEM have been proposed \cite{Skokos}.
\spi
The dynamic stability of the map, namely its sensitivity to small
random perturbations or to round-off, is another relevant issue.  The
Forward Error Method (FEM) consists in evaluating the divergence of
the perturbed orbit from the reference orbit. We do not consider
deterministic perturbations since an extended   literature exists at least
when the map is integrable or uniformly hyperbolic \cite{Arnold}.  To
analyze the round-off effects a convenient approach is based on the
Reversibility Error Method (REM), which consists in computing the
divergence of the initial point from its image after  $n$ forward and
$n$ backward iterations of the perturbed map, avoiding the
exact computation of the map.  This method is usually known as
\textit{Reversibility test} \cite{aaresath74} and it is routinely used
to analyze the regularization method for close encounters between two
massive objects \cite{reversTest}, or in the electromagnetic problems
\cite{chaosJ}, or to study the collisions in the few body
gravitational problem \cite{analRev}.  It has also been used to
investigate the dynamic stability of the Hamiltonian model
$H=p^2/2-1/x +\epsilon x \cos(\omega t)$ \cite{Casati}.
\spi
We might expect that, asymptotically, REM and FEM have a similar
behavior.   If the error is random we rigorously prove the
equivalence of REM and FEM asymptotic behavior in the case of linear symplectic
maps. To our knowledge this is a new result and can be extended to the
non linear case.  The extension of the proof to non linear maps
requires more sophisticated mathematical tools such as in
\cite{Oseledec}, and is not afforded here, even though we carry out the
basic preliminary steps.  Our numerical simulations suggest that this
equivalence holds also for non linear maps.  
For chaotic orbits, which have a positive maximum Lyapunov
characteristic exponent $\lambda>0$, the asymptotic divergence of REM,
FEM and LEM is governed by the same exponential law.  For regular
orbits, which have $\lambda=0$, the asymptotic divergence follows a
power law, with exponent $\beta$.  If the perturbation is random the
exponent for REM and FEM is $\beta=3/2$ for a generic observable (for
instance the separation) and for a generic anisochronous system.  If the
system is isochronous the exponent reduces to $\beta=1/2$.  If the
observable is a first integral the exponent is also $\beta=1/2$.
The Lyapounov error growth follows a power law with exponent $1$ for a
generic observable and for a generic anisochronous system.  For and
isochronous system or if the observable is a first integral the
exponent is $\beta=0$.  We provide a rigorous justification to the
above statements on the power law exponents for LEM, FEM and LEM and a
numerical check for the restricted planar circular $3$-body problem.
In the case of round-off the same power laws for REM are observed,
provided that the map, in the chosen coordinates system, is
sufficiently complex from the computational viewpoint.  \spi As a
counterpart of FEM we propose the Fidelity \cite{Marie}, which
measures the correlation between the unperturbed and perturbed orbits.
The Fidelity decay law is related to the asymptotic growth of the
global error.  A correspondence with REM is achieved by defining the
Fidelity as the correlation between an observable computed on the
initial point and its image after $n$ forward and $n$ backward
iterations of the perturbed map.  Rigorous results on the Fidelity for
prototype dynamical models with random perturbations are already known
\cite{Marie-Chaos}. The comparison of REM for round-off and noise was
considered in \cite{Mestre} and extended to the Fidelity in
\cite{Zanlungo_EL,Zanlungo}.
\spi
In the present paper we show that
for the planar circular three body problem REM, FEM and Fidelity are
suitable to investigate the effects of small random perturbations,
whereas REM and the corresponding Fidelity are suitable to explore the
round-off effects.  The $3$-body problem is the paradigm of systems
in which both regular and chaotic orbits coexist in a very narrow
region of the phase space and can provide a significant insight in
astrophysical relevant problems (see reference \cite{Valtonen}, for a
detailed discussion about the $3$-body problem and its application
in astrophysics). The results previously obtained on prototype
dynamical models are confirmed and extended.  Even though in the
neighborhood of the equilibria or periodic points in the rotating
system the Birkhoff normal forms can be used to approximate the
quasi-integrable dynamics with an integrable one (obtaining Nekhoroshev
stability estimates from bounds on the remainder), the numerical
integration procedure cannot be avoided to explore the whole phase
space in the Poincar\'e section.  
\spi 
The computation of REM on a
grid of points in phase space for a fixed number $n$ of iterations and
its visualization provides an easy insight on the dynamic stability
of a map.  Its use may be convenient to explore the boundary between
regions of regular and chaotic motion.  The Fidelity allows
to quantify the perturbation size and to determine the memory loss
rate of the orbits in a given invariant domain.  It  provides
information of statistical nature but it is also computationally more
demanding since a Monte-Carlo sampling of the invariant domain is
required. The REM and Fidelity are particularly suited to explore the
transition regions in complex dynamical systems with many degree of
freedom.  \spi The paper has 6 sections. Section \ref{sec1}:
introduction. Section \ref{sec2}: the $3$-body Hamiltonian in the fixed
frame, in the rotating frame and their symplectic integrators.
Section \ref{sec3}: analytic proof of the asymptotic equivalence of REM and
FEM, for a linear symplectic map with a stochastic perturbation.
Section \ref{sec4}: numerical analysis of REM for round-off, of REM and FEM for
random perturbations and comparison with LEM. Section \ref{sec5}: definition of
Fidelity and numerical results.  Section \ref{sec6}: conclusions.

%
%

\section{Three body Hamiltonian}
\label{sec2}
\def\dt{{\Delta t}}
\def\Tr{\,\hbox{Tr}\,}
The restricted  planar circular three body problem (see for instance \cite{Sz72} for a detailed description of this problem)
consists in a primary central body of mass $m_1$, a secondary body of
mass $m_2 \leq m_1$ describing circular orbits around their center of
mass, and a third body of mass $m_3$, so small that it does not
perturb the motion of the first two bodies.  We consider two reference
frames: the fixed inertial frame with the origin on the center of mass
and the rotating frame with the same origin and where the first two
bodies are at rest.  As customary we scale the space coordinates with
the distance $r_*$ between the massive bodies and the time with
$2\pi/T_*$ where $T_*$ is the period of the circular motion (see
\cite{turchetti} for details). Denoting with $t$ the scaled time and
$x_F,y_F$ the scaled coordinates of the third body in the fixed frame,
the Hamiltonian in the extended phase space, where we introduce a new
coordinate $\tau$ and its conjugate momentum $p_\tau$, reads

\begin{equation}\label{eq1}
   H_F   = T_F+V_F, \qquad \qquad T_F= {p_{x\,F}^2 + p_{y\,F}^2 \over 2} + p_\tau,  \qquad \qquad 
V_F  = -{1-\mu \over  r_1}-{\mu \over r_2},  
\qquad  \mu  = \frac{ m_2}{ m_1+m_2}      
\end{equation}
The potential is a periodic function of $\tau$ with period $2\pi$ and
$H_F$ is a first integral of motion.  Notice that $\tau$ is an angle,
its conjugate momentum $p_\tau$ is an action and that $t=\tau$.  In
the rotating frame the first and second bodies are on the $x$ axis
with coordinates $x_{1c}=-\mu$ and $x_{2c}=1-\mu$ and the distances
$r_1,r_2$ of the third body from the first two in the fixed frame are
given by
\begin{equation}\label{eq2}
r_1  = \sqrt{(x_F+\mu \cos \tau)^2+(y_F+ \mu \sin \tau)^2}  \quad 
    r_2  = \sqrt{(x_F-(1-\mu) \,\cos \tau )^2+(y_F-(1-\mu) \, \sin \tau )^2}  
\end{equation}		
assuming the rotation of the massive bodies with respect to the fixed frame is counter-clockwise.
The one period map is just the Poicar\'e section $\tau=0 \,\hbox{mod}\, 2\pi$ in the fixed frame.
\spi
In the rotating frame   the coordinates of the third body are $x,y$ and  the Hamiltonian is given by 
\begin{equation}\label{eq3}
  H = T+V \qquad \qquad T={p_x^2+p_y^2\over 2}+ yp_x-xp_y   \qquad \qquad 
 V=-{1-\mu \over r_1}-  {\mu\over r_2}     
\end{equation}
where the distances are now expressed by
\begin{equation}\label{eq4}
 r_1=  \sqrt{(x+\mu)^2+y^2},   \qquad \qquad 
r_2= \sqrt{(x-1+\mu)^2+y^2}  
\end{equation}
The Hamiltonian is conserved $H= -J/2 =E$
where $E$ is the energy and $J$ the Jacobi integral.
The first integral  $H$  can be written as  the sum  of the kinetic energy 
plus the effective potential,  sum of the gravitational and the centrifugal 
potentials 
\begin{equation}\label{eq5} 
 H= {{\dot x}^2+  {\dot y}^2\over 2} + V_{\hbix{eff}}(x,y) \qquad \qquad V_{\hbix{eff}}= V-\frac{x^2+y^2}{2}
\end{equation}
The coordinates of the Lagrange equilibrium points $L_4$ and $L_5$
(critical points of $V_{\hbix{eff}}$) are given by $x_c=1/2-\mu,\;
y_c=\pm \sqrt{3}/2$.  The massive bodies and $L_4$ or $L_5$ are the
vertices of an equilateral triangle.  We are interested in the
evolution of the system in the rotating frame: as a consequence may either
integrate the equations of motion of the Hamiltonian \ref{eq3}. As an
alternative, we transform the initial conditions to the fixed frame,
integrate the equations of motion of the Hamiltonian \ref{eq1},  and transform
back to the rotating frame whenever it is needed.  The last one is the
procedure we adopt.
\spa
We choose $\mu=0.000954$ which corresponds
approximately to the Jupiter-Sun masses and fix the Jacobi constant
$J=3.07$ close to the value
$J_c=-2V_{\hbix{eff}}(x_c,y_c)=3-\mu+\mu^2$ assumed at the equilibrium
points $L_4,\,\,L_5$ \cite{MD99}.  Our numerical analysis is referred
to a 3D manifold $\cal{M}_J$ in the 4D phase space, specified by a
given value of the Jacobi constant $J$.  We consider the 2D manifold
$\cal{M}_P=\cal{M}_J\cap \cal{L}$ obtained by intersecting the Jacobi
manifold with  the linear manifold $\Lcal:\{y=0$ and $\dot y>0\}$.
The projection of $\Mcal_P$ into the $(x,\dot x)$ phase plane is a domain
defined by
\begin{equation}\label{eq6}
{\dot x}^2 \leq x^2+ 2\, {1-\mu\over |x+\mu|}+ 2\,{\mu \over |x-1+\mu|} -J   
\end{equation}
The initial conditions are chosen in $\Mcal_P$ and we examine the
intersections of the orbit with $\Mcal_P$.  Their projections on the
$(x,\dot x)$ or $(x,p_x)$ phase planes are considered for
visualization. For each orbit the initial conditions are $x(0)=x_0$,
$y(0)=0$ and $\dot{x}(0)=v_{x\,0}$ chosen so that the inequality \ref{eq6}
is satisfied. The remaining initial condition $\dot{y}(0)$ is then
given by
\begin{equation}\label{eq7}
 \dot y(0)=  \sqrt{ x_0^2 -{v_{x\,0}}^2 + 2\, {1-\mu\over |x_0+\mu|}+ 2\,{\mu \over |x_0-1+\mu|}-J }
\end{equation}
\spa
    {\bf Symplectic integrator maps and errors}
\spa
We consider the
fourth order symplectic and symmetric integrator for the evolution, in the fixed and
rotating frame, generated by the Hamiltonians \ref{eq1} and \ref{eq3}.  The splitting of the Hamiltonian into two integrable components allows to introduce
a second order symplectic and symmetric evolution operator. By three compositions of the second
order operator, the fourth order symplectic and symmetric operator is obtained
\cite{Yoshida}.  The Lie derivative for the time independent
Hamiltonian $H$ is denoted by $D_H$ and the corresponding evolution
operator in a time interval $t$ is the Lie series $e^{t\,D_H}$.  In
the fixed frame we split the Hamiltonian $H_F$ according to equation
\ref{eq1} and  the evolution generated by $T_F$ and $V_F$ can be exactly computed.
The symmetric second order
scheme  is  defined by 
\begin{equation}\label{eq8}
 \Mop^{(2)}_{\dt}\equiv  e^{\Delta t/2\,\,D_{V_F}}\,\, e^{\Delta t\,\,D_{T_F}}   \,\,e^{\Delta t/2\,\,D_{V_F}} 
\end{equation}
the operator $\Mop^{(2)}_{\dt}$ advances the phase space vector $\xbf_F=(x_F, y_F,\tau, p_{xF},
p_{yF},p_\tau)$ from time $t$ to time $t+\Delta t$ with an error of
order $(\Delta t)^3$. The fourth order scheme is defined by
\begin{equation}\label{eq9}
\Mop^{(4)}_{\dt}= \Mop^{(2)}_{\alpha \Delta t}\,\, \Mop^{(2)}_{\beta \Delta t}\,\,\Mop^{(2)}_{\alpha \Delta t},
\qquad \qquad \alpha= {1\over 2 -2^{1/3}}, \qquad \beta= 1-2\alpha  <0.  
\end{equation}
The triple composition of the maps corresponding to the second order
evolution generates a symplectic map which advances $\xbf_F$ from $t$
to $t+\Delta t$ with an error $O(\Delta t)^5$.  \spa The procedure to
obtain the symplectic integrators in the rotating frame is the
same. With the splitting of the Hamiltonian according to equation \ref{eq3}
the symmetric  second order scheme in this case is given by 
\begin{equation}\label{eq10}
 \Mop^{(2)}_{\Delta t}= e^{\Delta t/2\,\,D_{V}}\,\, e^{\Delta t\,D_{T}}   \,\,e^{\Delta t/2\,\,D_{V}} 
\end{equation}
the new operator $\Mop^{(2)}_{\Delta t}$ advances the phase space vector $\xbf=(x,y,p_x,p_y)$ from time
$t$ to time $t+\Delta t$ with an  error of order $(\Delta t)^3$. The
fourth order scheme is defined by \ref{eq9} as in the previous case.  Higher
order schemes are very easily obtained.  For instance the sixth order
scheme is given by equation \ref{eq9} where $M^{(4)}$ and $M^{(2)}$ are replaced by
$M_6$ and $M_4$ with $\alpha= 1/(2-2^{1/5})$. The eight order scheme
is given by equation \ref{eq9} where $M_4$ and $M_2$ are replaced by $\Mop^{(8)}$ and
$\Mop^{(6)}$ with $\alpha= 1/(2-2^{1/7})$.
\spa
The number of evaluations
of $\Mop^{(2)}$ for integrators of order $2m$ grows as $3^{m-1}$
(optimized algorithms lower this number to $7$ for $m=3$ and to $15$
for $m=4$ see \cite{Yoshida}). The local error of order
$(\dt)^{2m+1}$ introduces fluctuations $\Delta H/H$ of order $(\Delta
t)^{2m}$   along the orbit.  Even though high order integrators seem
to be convenient, the appearance of numerical instabilities for large
$m$ suggest the choice $m=2$ as a reasonable compromise for our
numerical exploration of the effects of round-off and random
perturbations.  \spa To any evolution operator $\Mop$ corresponds a
map $M$, to operators multiplication corresponds the composition of
maps.  To $\Mop^{(2)}_{\dt}$ and its inverse ${ \Mop^{(2)}_{\dt}} ^{-1}$ we
associate the maps $M^{(2)}_{\dt}$ and ${M^{(2)}_{\dt}}^{-1}$ and we denote
$M^{(2)}_{\eps,\,\dt}$ and ${M^{(2)}_{\eps,\,\dt}}^{-1}$ the
corresponding maps with a round-off or random perturbation of
amplitude $\epsilon$. The perturbed inverse
${M^{(2)}_{\eps,\,\dt}}^{-1}$ is not the inverse of the perturbed map
so that
\begin{equation}\label{eq11}
 \bigl ( {M^{(2)}_{\eps,\,\dt}}^{-1}\circ M^{(2)}_{\eps,\,\dt}\bigr)  (\xbf)\not=\xbf   
\end{equation}
The fourth order perturbed map $M^{(4)}_{\eps,\,\dt}$, obtained as the
composition of three second order perturbed maps, is also
irreversible.  The forward error $M^n_\epsilon(\xbf_0)-M^n(\xbf_0)$
and the reversibility error $M^{-n}_\epsilon\circ M^n_\epsilon(\xbf_0)
-\xbf_0$ can be analyzed for different choices of the symplectic map $M$. 
In the scaled variables the period is $T=2\pi$ and the time step we choose is
$\dt=T/n_s$, where $n_s$ is an integer.  \spi In the next Section we
choose $M$ to be the one period map $M= \left( {M^{(4)}_{\dt}}\right)^{n_s}$ obtained from
the Hamiltonian $H_F$ in the fixed frame.  As a consequence the local error on the map $M_{\eps}$ is
the global error of $M^{(4)}_{\eps,\,\dt}$ after $n_s$ iterations.
\\ In Section \ref{sec4} to compute the Fidelity we choose $M$ to be the
Poincar\'e map in the rotating frame, as before the map is computed using the
symplectic integrator in the fixed frame.  The intersection with the
hyperplane $y=0$ can be computed using linear interpolation if the
stochastic perturbation of the map is large with respect to the
interpolation error ($\sim 10^{-10}$). An interpolation to machine
accuracy is provided by the H\'enon method \cite{Henon} but its
application is straightforward only if $M^{(4)}_{\dt}$ is the symplectic
integrator in the rotating frame.

%
%

%
%

%
%
\spazio
\section{Asymptotics of  Lyapunov, Forward and Reversibility Errors}
\label{sec3}
\def\eps{\epsilon}
In this section we analyze the asymptotic behavior of the forward and
reversibility errors.  Even though we may expect that the behavior is
the same a mathematical proof is necessary to make  this expectation
a solid statement.  Assuming the amplitude $\eps$ of the local perturbation is infinitesimal, the
general expression for the global forward and reversibility errors are
obtained at first order in $\eps$.  Explicit asymptotic expressions are given for random
perturbations in the case of linear maps.  As a consequence the power
law growth of the global error for an integrable map and the
exponential growth for an expanding (chaotic) map are easily
recovered.  Extending the proof to non linear maps requires a more
sophisticated mathematical apparatus.
\subsection{Lyapunov error}
Let us consider a symplectic map $M(\xbf)$ and its orbit
$\xbf_n=M(\xbf_{n-1})$ with initial point $\xbf_0$.  Consider a nearby
point $\xbf_0 + \eps \,\ebf$, as initial condition for another orbit,
where $\ebf$ is a vector of norm 1 and $\epsilon$ is a small
parameter. The Lyapunov error is defined as the distance of these
orbits after $n$ iterations
\begin{equation}\label{eq12}
d^{(L)}_n= \Vert  M^n(\xbf_0 + \eps \,\ebf) -M^n(\xbf_0) \Vert .
\end{equation}
 The asymptotic limit of this error, defined by
\begin{equation}\label{eq13}
\lambda= \lim_{n\to \infty} \lim_{\eps\to 0} \,\ln{\left( d^{(L)}_n\over \epsilon \right) },   
\end{equation}
gives, for almost all the directions $\ebf$ (namely for all the points
on the unit sphere $\Vert \ebf \Vert =1$ except for a set of measure
zero) the maximum Lyapunov exponent. If the system is ergodic, or if
we consider an ergodic component,  the limit is the same for almost all
initial conditions $\xbf_0$.  If we consider the parallelepiped (parallelotopes)
$\Pop_k$, whose sides are $\epsilon \,\ebf_i$ for $i=1,2,\ldots,k$ and
$k$ ranges from $2$ up to the phase space dimension $d$, supposing
$\ebf_i$ are linearly independent vectors, the asymptotic behavior of
the volumes of $\Pop_k$ determines the Lyapunov spectrum, see
\cite{Oseledec,Benettin}.
\subsection{Forward error}
Rather than considering, for a given map, the error due to an initial
displacement to analyze the sensitivity to initial conditions, we may
consider a small perturbation of the map to explore the sensitivity to
small changes of the laws of motion.  We denote with
$M_\epsilon(\xbf)$ the perturbed map, where the perturbation is due to
round-off or random errors.  As explained in the introduction we do
not consider small deterministic perturbations due to abundant
literature on the subject. The orbit of the perturbed map is defined
by
\begin{equation}\label{eq14}
\xbf_{\eps,\,n}=M_\eps(\xbf_{\eps,\,n-1}) = M(\xbf_{\eps,\,n-1}) + \epsilon\xibf_n,  \qquad   \qquad   n\geq 1,  
\end{equation}
The initial point $\xbf_{\eps,\,0}= \xbf_0+\epsilon \xibf_0$ can be
perturbed, but we shall assume it is not, choosing $\xibf_0=0$.  With
$\epsilon$ we denote the perturbation amplitude.  For a given
round-off error of amplitude $\epsilon$ the exact map $M$ can only be
approximated by using a higher accuracy (where the round-off
error is typically $\epsilon^2$).  The round-off error is defined by
$\epsilon \xibf_n= M_\eps(\xbf_{\eps,\,n-1}) -M(\xbf_{\eps,\,n-1}) $.
In the case of a random perturbation we may evaluate $M$ with the
selected machine accuracy provided that $\eps$ is larger by some
orders of magnitude with respect to the round-off. The $\xibf_n$ are
independent random vectors whose components have zero mean and unit
variance.
The  global error at step $n$ is defined by
\begin{equation}\label{eq15}
\eps \Xibf_n=\xbf_{\eps,\, n} - \xbf_ n = M^n_\eps(\xbf_0)-M^n(\xbf_0).  
\end{equation}
and the forward error (FEM) is defined as the mean squares deviation namely
\begin{equation}\label{eq16}
d_n= \langle \Vert M_\eps^n(\xbf_0) -M^n(\xbf_0) \Vert ^2 \rangle^{1/2}  
\end{equation}
where $\langle \;\; \rangle$ denotes the average over the stochastic
process. If the round-off is considered, then the forward error is
defined just by the distance (no average). The global error due to round-off 
is similar to the one due to a random perturbation, if the map
has a sufficient computational complexity. However we have access only
to a single realization corresponding to the hardware we use.  
\spi
The global error is related to the local errors according to
\begin{equation}\label{eq17}
\begin{aligned}
\eps {\bm \Xi}_n   = & M_\eps({\bm x}_{\eps, n-1} )  - M({\bm x}_{\eps, n-1} ) \quad +\quad  M({\bm x}_{\eps,n-1})-M({\bm x}_{n-1})
\\
 = &  \eps {\bm \xi}_n + \eps DM({\bm x}_{n-1}) {\bm \Xi}_{n-1} +  O(\eps^2) = \\
 = &  \eps {\bm \xi}_n + \eps DM({\bm x}_{n-1}){\bm \xi}_{n-1} + 
\eps DM( {\bm x}_{n-1}) DM( {\bm x}_{n-2}){\bm \Xi}_{n-2} + O(\eps^2)   \\      
\end{aligned}
\end{equation}

\spa
at first order in $\eps$, where $DM$ is the tangent map, namely $(DM)_{i,j}=\partial M_i/\partial x_j$. 
Recalling that $DM^2( {\bm x}_{n-2})= DM( {\bm x}_{n-1})  DM({\bm x}_{n-2})$  the final
result  reads 
\begin{equation}\label{eq18}
 \eps {\bm \Xi}_n = \eps \sum_{k=1}^n DM^{n-k}({\bm x}_k) \xibf_k +
  O(\eps^2)  
\end{equation}
When the initial error is not zero  an  additional term  $M^n(\xbf_0+\eps \xibf_0) -M^n(\xbf_0) $
must be included and equation \ref{eq18} still holds with the sum starting from $k=0$.
\subsection{Reversibility error}
The reversibility error is given by the distance of the initial point from the point obtained 
iterating it $n$ times forward and $n$ times
backward with the perturbed map. For the unperturbed map this error
vanishes since the map is reversible.  We denote with $M^{-1}(\xbf)$
the inverse map and with $M_\eps^{-1}(\xbf)$ the perturbation of the
inverse map. The local error at iteration $n$ is denoted by
$\xibf_{-n}$ and
\begin{equation}\label{eq19}
 \xbf_{\eps,\,-n}=M^{-1}_\eps(\xbf_{\eps,\,-n+1}) = M^{-1}(\xbf_{\eps,\,-n+1}) + \epsilon\xibf_{-n } \qquad   \qquad   n\geq 1
\end{equation}
Notice that the perturbed inverse map differs from the inverse of the
perturbed map namely $M_\eps^{-1}\circ M_\eps (\xbf_0) \not = \xbf_0$.  Indeed we have
\begin{equation}\label{eq20}
\begin{aligned}
M^{-1}_\eps(M_\eps(\xbf_0)) & =  M^{-1}_\eps(\xbf_1+\eps \xibf_1) = \\
& = M^{-1}(M(\xbf_0) +\eps \xibf_1)+ \eps \xibf_{-1}=
\xbf_0+ \eps DM^{-1}(\xbf_1)\,\xibf_1 + \eps \xibf_{-1} +O(\eps^2) 
\end{aligned}
\end{equation}
More generally we define the global reversibility error according to
\begin{equation}\label{eq21}
 \epsilon \Xibf_n^{(R)}= M^{-n}_\eps \circ M^n_\eps(\xbf_0)- \xbf_0  
\end{equation}
In order to evaluate the global error at the first order in $\eps$ we
start with a recurrence that can be proven by induction.  At the first
step we have
\begin{equation}\label{eq22}
  M^{-1}_\eps M^n_\eps(\xbf_0) = M^{-1}(\xbf_n+ \eps \Xibf_n)+ \eps \xibf_{-1} =
  \xbf_{n-1} +\eps DM^{-1}(\xbf_n) \Xibf_n+ \eps \xibf_{-1}  +O(\eps^2)  
\end{equation}	
Then after $m$ iteration of the perturbed inverse map we obtain 
\begin{equation}\label{eq23}
\begin{aligned}
M_\eps^{-m}\circ M_\eps^n({\bm x_0}) = \xbf_{n-m} +\eps DM^{-m}({\bm x}_n) {\bm \Xi}_n + \eps \sum_{k=1}^m DM^{-(m-k)} ({\bm x}_{n-k}) \xibf_{-k}\, +O(\eps^2),     
\end{aligned}
\end{equation}
Setting $m=n$ in the previous relation we obtain the expression of the
reversibility error

\begin{equation}\label{eq24}
\begin{aligned}
\eps {\bm \Xi}^{(R)}_n & = M_\eps^{-n}M_\eps^n ({\bm x}_0) -{\bm x}_0= \\
& = \eps DM^{-n}(  \xbf_n   )\,{\bm \Xi}_n + \eps \sum_{k=1}^n
DM^{-(n-k)}({\bm x}_{n-k})  \xibf_{-k} +O(\eps^2).      
\end{aligned}
\end{equation}
We compare the growth with $n$ of the forward error   
$d_n= \epsilon \langle \Vert {\bm \Xi}_{n}\Vert^2 \rangle ^{1/2} $
  with the reversibility error   $d_n^{(R)}= \epsilon \langle \Vert
 {\bm \Xi}^{(R)}_{n}\Vert ^2 \rangle ^{1/2} $.   The distance
 $d_n^{(R)}$ vanishes for the unperturbed map.  The perturbation
 is a random vector $\xibf$ with independent components of unit
 variance $\langle (\,{\bm \xibf}_k)_j({\bm
   \xibf}_{k'})_{j'} \rangle =\delta_{kk'}\delta_{jj'}$ if $kk'>0$. The average vanishes
   if $kk'<0$, since the perturbation of the map $M$ and the perturbation of its inverse are independent.
   Taking into account that
 $\langle A\xibf\cdot A\xibf \rangle = \Tr(\,A\,A^T)$ for any matrix
 $A$ the result for $d_n^2$   is
\begin{equation}\label{eq25}
 d_n^2= \epsilon^2\, \langle \Vert {\bm \Xi}_{n}\Vert^2  \rangle    = \eps^2 \sum_{k=1}^n \Tr \left[DM^{n-k}({\bm x}_k)
 \left( DM^{n-k}({\bm x}_k)\right)^T \right]     +O(\epsilon^3) 
\end{equation}
 where the suffix $^T$ denotes the matrix transpose. The result
 for   $\bigl(\,d^{(R)}_n\bigr)^2 $   is  
\begin{equation}\label{eq26}
\begin{aligned}
  \left (   d_n^{(R)} \right )^2=\epsilon^2
\left \langle  \Vert {\bm \Xi}^{(R)}_{n}\Vert ^2   \right \rangle =   
&  \eps^2 \sum_{k=1}^n \Tr\left[ DM^{-n}({\bm x}_n) DM^{n-k}({\bm x}_k) \left( DM^{-n} ({\bm x}_n )
    DM^{n-k}({\bm x}_k) \right)^T + \right .  \\
& \left .  +  DM^{-(n-k) }({\bm x}_{n-k})     \left(DM^{-(n-k) }({\bm x}_{n-k})\right)^T \right ]   
  +O(\epsilon^3)   .      
\end{aligned}
\end{equation}
We now show that the growth of $d_{n}$ and $d_n^{(R)}$ is
comparable. This can be easily proved if $DM(\xbf)=A$ is a constant
symplectic matrix. In this case we have
\begin{equation}\label{eq27}
\begin{aligned}
  d_n^2 
& =\eps^2 \sum_{k=0}^{n-1} \Tr \Bigl [ A^{k} \left(A^{k}\right)^T \Bigr] \,+\,O(\epsilon^3), \\ 
 \left (d_n^{(R)}\right)^2 
&= 2\, \eps^2 \sum_{k=0}^{n-1}  \Tr\, \Bigl[ A^{-k} \left(A^{-k} \right)^T \bigr] \quad  +\quad 
\eps^2 \Tr\, \Bigl[ A^{-n} \left(   A^{-n}   \right)^T  -I \bigr]        \,+\,O(\epsilon^3).              
\end{aligned}
\end{equation}
We recall that if $A$ is a real symplectic matrix its inverse $A^{-1}$
has the same eigenvalues.
%
%
In addition when the multiplicity is higher than $1$ the Jordan
form must be considered.
Supposing that all the eigenvalues are simple and that $e^\lambda$ is
the largest eigenvalue (or the largest modulus in the complex case) where
$\lambda>0$, we have
\begin{equation}\label{eq28}
\lim_{n\to \infty }\, \lim_{\epsilon \to 0 }\,\,{1\over n}\,\ln \,\parton {d_n\over \epsilon}  = \lim_{n\to \infty } \,\lim_{\epsilon \to 0 }\, \,{1\over n}\,\ln\,\parton {d_n^{(R) }\over \epsilon}= \lambda,  
\end{equation}
If the eigenvalues of $A$ have unit modulus  then the above limit is zero. In this case   the asymptotic 
growth of  $d_n$ and $d_n^{(R)}$ follows a power law. 
More specifically if the eigenvalues of $A$ are real and $e^{\lambda }$ with $\lambda >0$ is the largest one, 
the  error growth is given by   $d^{(R)}_n\sim \sqrt{2}\, d_n\sim c\,\epsilon\, e^{\lambda n }$. 
\pan 
If all the eigenvalues are complex with unit  modulus then $ d^{(R)}_n\sim \sqrt{2}\, d_n\sim c \,\epsilon\, n^{1/2}$. 
If  $A$ is reducible to a Jordan form whose blocks have the form  
$ A= \left (\begin{array} {cc} 1 &  \alpha   \cr    0 & 1  \cr  \end{array} \right ) $
then $d^{(R)}_n\sim \sqrt{2}\, d_n\sim c \,\epsilon\,\alpha \,n^{3/2}$.  In general for an integrable system the 
error  growth is given by  
\begin{equation}\label{eq29}
 d^{(R)}_n \sim  c\,\epsilon\,\parton{ n +{\alpha^2}\,  {n^3\over 3}}^{1/2}
\end{equation}
The system is isochronous when $\alpha=0$. The anisochronous character of a system   in numerical
simulations emerges with the $n^{3/2}$ asymptotic behavior only  if  $\alpha$  is above  a threshold. 
A least squares fit of the form $c n^\gamma $ 
provides  a value for $\gamma$ which smoothly varies between $ 1/2$ and $3/2$ with a transition occurring 
for $\alpha\sim 1$.  See \ref{appendix-sec1} for more details. 
%

%
%

%
%

\section{Numerical analysis of  global errors growth}
\label{sec4}
\begin{figure}[!ht]
\centering
 \includegraphics[width=7cm,  height=7cm]{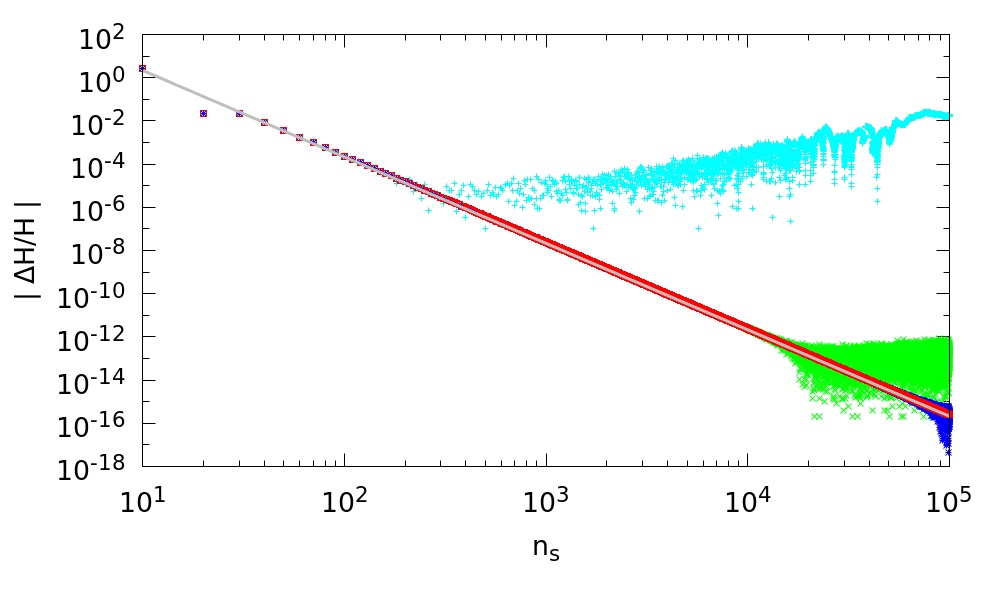}
 \includegraphics[width=7cm,  height=7cm]{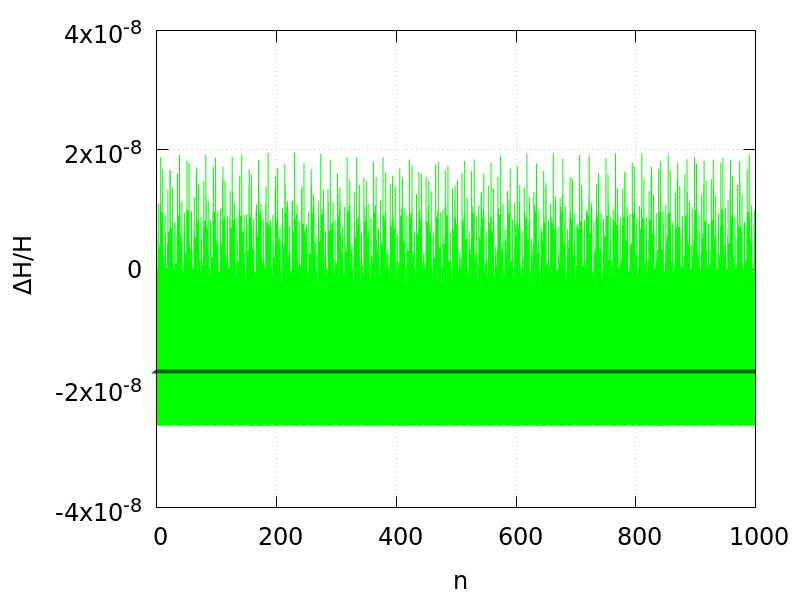}
 \caption{\textit{Integration error and variation of $H$}.
  Left: variation of the first integral
  $H$ (in the rotating frame Hamiltonian) $|\Delta H/H|$ in one period
  $T$, as a function of the number of integration steps per period
  $n_s$.  Results for different machine precision are compared: single
  (cyan), double (green), extended (blue) and quadruple precision
  (red). The integrations are based on a fourth order symplectic scheme
  for the Hamiltonian $H_F$, see equation \ref{eq1}, with time step $\Delta
  t=T/n_s$.  The error saturates when the machine accuracy is reached.
  The gray line is the linear ln-ln fit $|\Delta H/H|=c\,
  n_s^{-\gamma}$ where $\gamma=-4$ within the statistical errors.  The
  initial conditions are chosen for a regular orbit, with initial
  conditions $x(0)=0.55$, $y(0)=0$, $\dot{x}(0)=0$ and $\dot{y}(0)$
  defined by the value of the Jacobi constant set up equal to
  $J=3.07$.  Right: variation of $| \Delta H |$ along the orbit $M^n(\xbf_0)$.
  Even though $\Delta H$ fluctuates, its average vanishes.}
\label{fig:1}
\end{figure}

\begin{figure}[!ht]
\centering
\includegraphics[width=8 cm, height=8 cm]{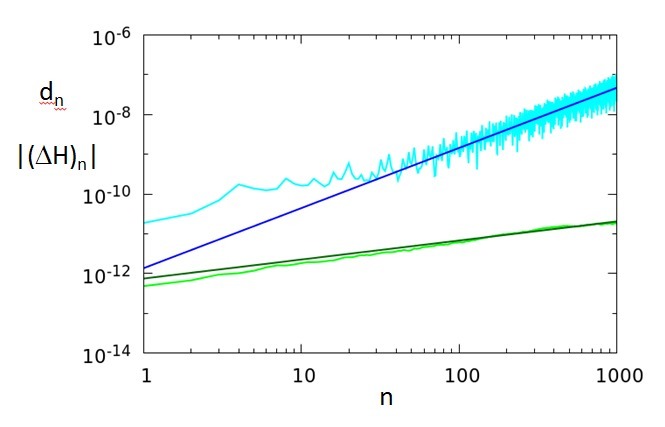}
\includegraphics[width=8 cm, height=8 cm]{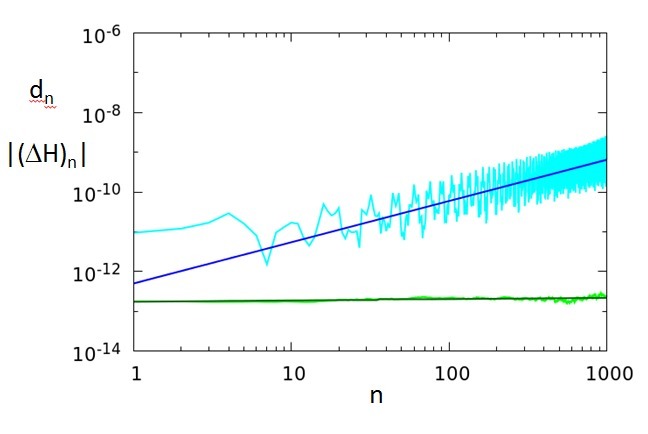}
\caption{\textit{ FEM and LEM errors and power law fit for a regular
    orbit.} Left panel: evolution of the Forward Error $d_n$ (cyan) and
  $(\Delta H)_n $ (green) due to a random perturbation of amplitude
  $\epsilon=10^{-13}$ with $100$ realizations of the noise for the
  regular orbit with initial condition $x(0)=0.55$. The straight lines are the least squares fit
  with $\beta_=1.51 \pm 0.05$ (blue) and $\beta_H=0.48\pm 0.01$ (dark-green).  Right
  panel: evolution of the Lyapunov Error $d_n$ and $(\Delta H)_n $
  with a perturbation to the initial condition $x(0)+\epsilon$ with
  $\epsilon=10^{-13}$. The straight lines correspond to the least
  squares fits with $\beta_d =1.04 \pm 0.07,\,\, \beta_H=0.03 \pm
  0.01 $.  This low value of $\beta_H$, though not zero (three
  standard deviations are required to reach it), is explained as
  the round-off effect which start to be appreciable precisely
  around $n=1000$.  The interval where the fit is computed is
  $50\leq n\leq 1000$.  The  round-off error  rises as $10^{-15} (nT)^ {1/2}$ and
for $n=1000$  becomes  appreciable. } \label{fig:3}
\end{figure}

\def\dt{{\Delta t}}
\def\eps{{\,\epsilon}}
Let $M_{\dt}$ be   the   symplectic integrator map   
for a time step $\dt=T/n_s$  and $ M_{\eps,\dt} $ be
the perturbed map.  We consider the one period  map 
$M=M^{n_s}_{\dt}$ and its perturbation $M_\epsilon=M^{n_s}_{\eps,\dt}$.
The  discretization error of the one period map  can be estimated
by the variation of the first integral of motion ($H_F$ or $H$).
For a fourth order integrator the discretization error  scales as $n_s^{-4}$
and saturates when the machine accuracy is reached,   as shown by figure
\ref{fig:1} left. 
If we choose $n_s$ below this threshold then the value of the first
integral along the orbit of the map oscillates without growing. A
power law fit to the growth of $(\Delta H)_n=|H\left(M^n(\xbf_0)
-H(\xbf_0)\right)|$ with the number of periods $(\Delta
H)_n=C\,n^{\beta}$  gives $\beta=0$ within the numerical
uncertainties, see figure \ref{fig:1} right. This is true as long as the
error growth due to the round-off is negligible. 
\spi
To evaluate the dynamic stability of the map we
consider its perturbation $M_\epsilon$ and look at two different type
of errors: the distance $d_n$ of the perturbed orbit from the
reference one and the variation $(\Delta H)_n$ of the first integral
along the perturbed orbit. We compare this error with the
reversibility error $d^{(R)}_n$ and $(\Delta H)_n^{(R)}$. In the case
of round-off we have access only to the reversibility error.  In the
case of random errors the numerical simulations support the asymptotic
equivalence of the forward and reversibility errors, which in the
previous section was proved to hold for linear symplectic maps. In addition the
asymptotic behavior of reversibility errors looks very similar for
round-off and random perturbations, when a single realization is
considered.  In the case of random errors a smooth behavior is
obtained averaging over many realizations and a good agreement with
the theoretical predictions is obtained.  For the round-off different
hardwares give different results with variations very similar to the
ones obtained for different realizations of random perturbations.  The
reversibility error for $n=1$ (the application of the one period map
and its inverse) due to round-off is almost
independent on the number $n_s$ of steps per period, in a wide range
$10\leq n_s\leq 10^4$, and it is close to the machine accuracy. This is due to the use of 
symmetric and symplectic integrator.
\spi
The numerical analysis we present
refers to the map which integrates the Hamiltonian $H_F$ in the fixed
frame, choosing the mass ratio $\mu=0.000954$ (close to the
Jupiter-Sun case). The initial conditions are chosen in the rotating
frame for the same value of the Jacobi constant $J=3.07$ (on the
Lagrange  point $L_4$ we have $J=J_c\equiv 2.9990468$).  We choose
$y(0)=\dot x(0)=0$ and $x(0)=0.55$, as initial conditions for a
regular orbit and $x(0)=0.56$ for a chaotic orbit.  The value of $\dot
y(0)>0$ is fixed by equation \ref{eq7}. In \ref{fig:6} we show the phase
portrait of these orbits and nearby ones on the Poincar\'e manifold
$\Mcal_P$ projected on the $\left( x,\dot x \right)$ phase plane.
\begin{figure}[!ht]
\centering
\includegraphics[width=8 cm, height=8 cm]{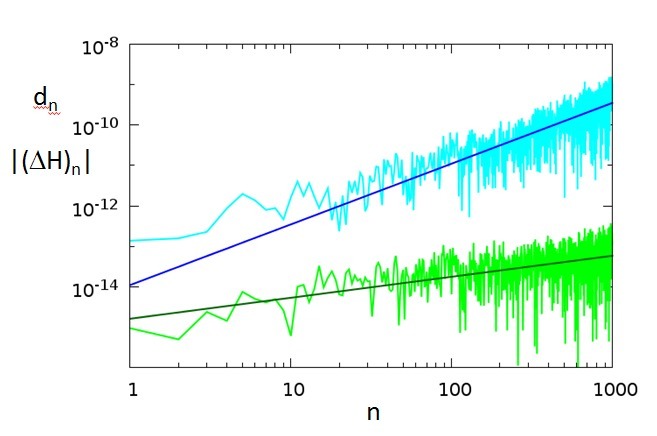}
\includegraphics[width=8 cm, height=8 cm]{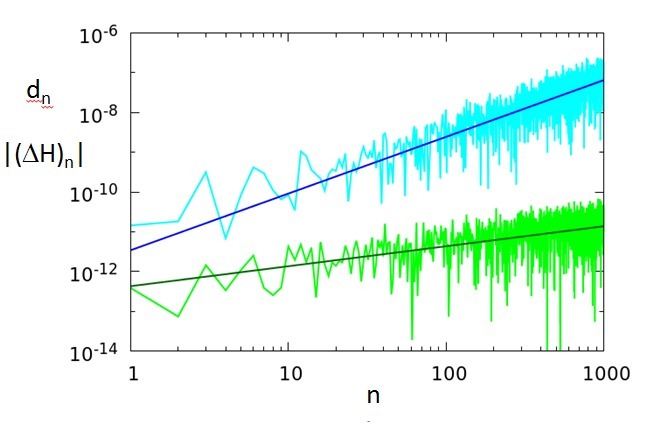}
\caption{\textit{REM errors and power law fit for a regular orbit}
  Left panel: evolution of the Reversibility Error $d_n^{(R)}$ (cyan) and
  $(\Delta H)^{(R)}_n$ (green) due to the round-off computed for a regular orbit with
  initial condition $x(0)=0.55$ and $y(0)=\dot x(0)=0$ the value of $\dot
y(0)>0$ is fixed by equation \ref{eq7}. The straight blue line is the least squares fit to
  $d_n^{(R)}= Cn^{\beta_d}$. The dark-green line is the least squares fit to $(\Delta H)_n^{(R)}= Cn^{\beta_H}$.
  The exponents are $\beta_d=1.50 \pm 0.09, \,\, \beta_H=0.52\pm 0.1$.  Right
  panel: evolution of the reversibility error for a stochastic
  perturbation of amplitude $\epsilon=10^{-13}$.  The straight lines
  correspond to the least squares fit with $\beta_d=1.43\pm 0.09,
  \,\, \beta_H=0.5\pm 0.1 $.  The interval where the fit is computed
  is $50\leq n\leq 1000$. }
\label{fig:2} 
\end{figure}
In figure \ref{fig:2} we show, for the regular orbit, the plot of the REM errors
$d^{\,(R)}_n$ and $(\Delta H)_n^{(R)}$ for the round-off (left panel) and  for a random 
perturbation (right panel). The computations
are in double precision so that the round-off error amplitude is
$\epsilon\sim 10^{-16}$ whereas the random perturbation amplitude is
$\epsilon= 10^{-13}$.  In all the figures the time step is fixed to $n_s=1000$.
In figure \ref{fig:3}, for the same orbit, we show the
plot of forward global errors FEM $d_n$ and $(\Delta H)_n$ for the
same random perturbation, after averaging on $100$ realizations (left panel) and the
Lyapunov global error LEM (right panel). The initial condition for LEM is varied
according to $x(0)+\epsilon$ with $\epsilon=10^{-13}$.  The error
growth follows a power law
\begin{equation}\label{eq30}
  d_n  \sim \epsilon  \,n^{\beta_d} \qquad \qquad 
  (\Delta H)_n  \sim \epsilon  \,n^{\beta_H}  
\end{equation}
For an integrable map the theoretical prediction for REM and FEM
errors due to a random perturbation is $\beta_d=3/2$ and
$\beta_H=1/2$, whereas $\beta_d=1$ and $\beta_H=0$ for the Lyapunov
error. The straight lines in the figures are obtained by least
squares fits. In the table 1 we quote the corresponding
values of the exponents $\beta_d $ and $\beta_H$.
For the round-off the value of $\beta_d $ is compatible  
$3/2$ and the value of $\beta_H$ is compatible with $1/2$,
which are theoretically predicted for the random perturbations.
The variations of the exponents  with different hardware implementations
of the round-off  are  similar  to  the changes  observed between 
different realizations of random perturbations.
\begin{figure}[!ht]
\centering
\includegraphics[width=8 cm, height=8 cm]{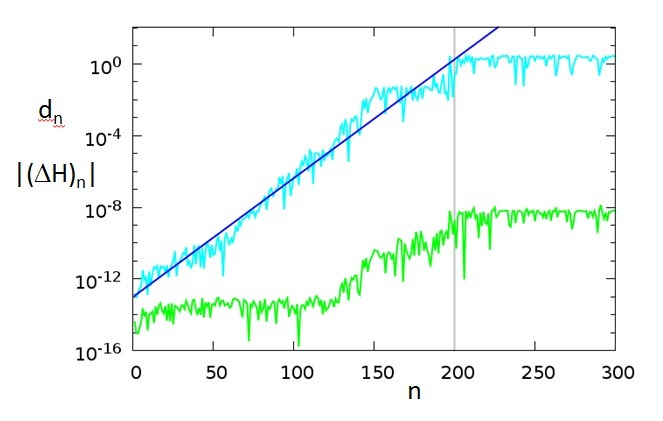}
\includegraphics[width=8 cm, height=8 cm]{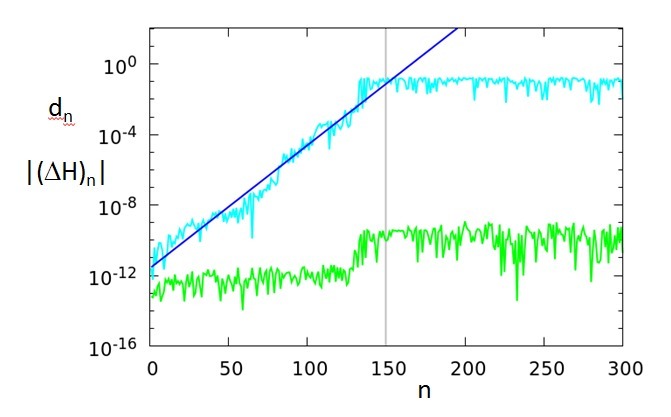}
\caption{\textit{REM errors and power law fit for a chaotic orbit.}
  Left panel: evolution of the Reversibility Error $d_n$ (cyan) and $(\Delta
  H)_n$ (green) due to round-off computed for a chaotic orbit with initial condition $x(0)=0.56$. The
  straight line corresponds the least squares fit with
  $\beta_d=0.0067\pm 0.002$.  Right panel: evolution of the
  reversibility error for a stochastic perturbation of amplitude
  $\epsilon=10^{-13}$.  The straight line corresponds to the least
  squares fit with $\beta_d=0.071\pm 0.003$.  The fitting interval
  $1\leq n \leq 200$ for the left panel, $1\leq n\leq 150$ for the
  right panel.}
\label{fig:4} 
\end{figure}

\begin{figure}[!ht]
\centering
\includegraphics[width=8 cm, height=8 cm]{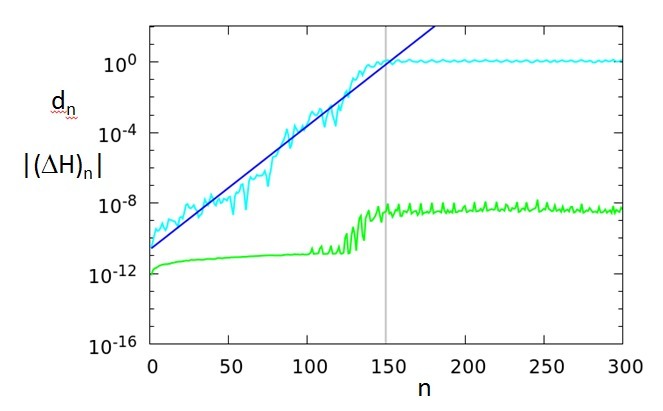}
\includegraphics[width=8 cm, height=8 cm]{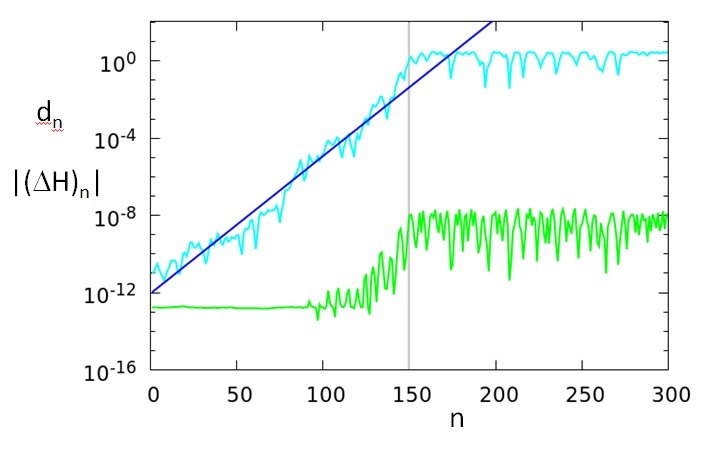}
\caption{\textit{ FEM and LEM errors and power law fit for a chaotic
    orbit.} Left panel: evolution of the Forward Error $d_n$ (cyan) and
  $(\Delta H)_n$  (green) due to a random perturbation of amplitude
  $\epsilon=10^{-13}$ with $100$ realizations of the noise for the
  chaotic orbit $x(0)=0.56$. The straight line is the least squares fit
  with $\beta_d=0.070\pm 0.002$.  Right panel evolution of the
  Lyapunov Error with a perturbation to the initial condition
  $x(0)+\epsilon$ where $\epsilon=10^{-13}$ The straight lines
  corresponds to the least squares fit with $\beta_d =0.071\pm 0.002
  $.  The fitting interval is  $1 \leq n \leq 150$.  }
\label{fig:5} 
\end{figure}

\spa
In figure \ref{fig:4} we show the plot of REM errors due to the round-off (left panel) and
to a random perturbation (right panel) for a chaotic orbit. In figure \ref{fig:5} the plot of
 the FEM error for a random perturbation (left panel) and LEM error (right panel) is shown for the
 same chaotic orbit. In this case the growth
 of the distance is exponential
\begin{equation}\label{eq31}
 d_n  \sim \epsilon \,10^{ \beta \,n }  \qquad \qquad  \lambda = {\beta\over T} \ln 10
\end{equation}
where $\lambda$ is the maximum Lyapunov exponent.  The orbits have been
computed up to $n=300$ periods. Notice that REM with round-off
saturates at $n\sim 200$ whereas REM and FEM errors for stochastic
perturbations and LEM saturate at $n\sim 150$. This is easily explained if we notice that 
in the first case $\epsilon\sim 10^{-16}$,
whereas in the second case $\epsilon\sim 10^{-13}$. The least squares fit gives $\beta_d=0.07$ which
corresponds to $\lambda\sim 0.0256$, in good  agreement with
the value obtained for the maximum Lyapunov exponent $\lambda$ computed with the renormalization method, which avoids
saturation.

Also in this case the REM  error due to  round-off and stochastic perturbations exhibit
the same behavior.  The error on the first integral $H$ remains constant
for a while, then has an exponential growth with about the same
coefficient $\beta_H=0.07$ and saturates to a value close  to the variation
$\Delta H$ of the first integral  $H$ along the unperturbed orbit due
to the truncation error. 
\spa
In the table 1 we resume the values of the exponents 
obtained by fitting the global error data for FEM, REM and LEM.
%
%
\spa
\begin{table}[!ht]
\centering
\label{table}
\begin{tabular}{|c|c|c|c|c|}
\hline
   Regular orbit       &     REM round-off     &     REM stochastic    &  FEM   stochastic     &     LEM                  \\ \hline
       $\beta_d$       &       $1.50\pm0.09$   &       $1.43\pm0.09$   &    $1.51\pm0.05$        & $1.04\pm0.07$        \\ \hline
       $\beta_H$       &       $0.52\pm0.1$    &   $0.5\pm0.1$         &     $0.48\pm0.01$       & $0.03\pm0.01$         \\ \hline
\end{tabular}
\spa
\begin{tabular}{|c|c|c|c|c|}
\hline
   Chaotic orbit       &     REM round-off     &     REM stochastic    &  FEM   stochastic     &     LEM                  \\ \hline
       $\beta_d$       &       $0.067\pm0.002$   &       $0.071\pm0.003$   &    $0.070\pm0.002$        & $0.071\pm0.002$        \\ \hline
\end{tabular}
\caption{\textit{Table of exponents.}  The exponents of the power law for regular
orbits and of the exponential law for chaotic orbits obtained by fitting
the simulation results for REM, FEM and LEM errors are presented.}
\end{table}
%
%

{\bf Phase space REM  plots }
\spazio
The geometry of orbits is usually inspected by considering the Poincar\'e map $M_P$
on the 2D manifold  $\Mcal_P$.  
The orbits are
visualized by projecting them into the $\left( x,\dot x \right)$ phase plane. In this case
the machine accuracy is not required for the intersection and a linear
interpolation is adequate.
\begin{figure}[!ht]
\centering
\includegraphics[width=6.9 cm, height=5.8 cm]{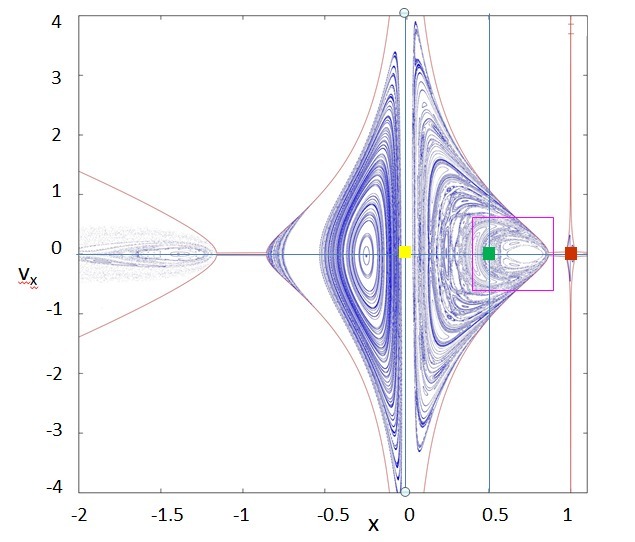}
\includegraphics[width=6.9 cm, height=6.0 cm]{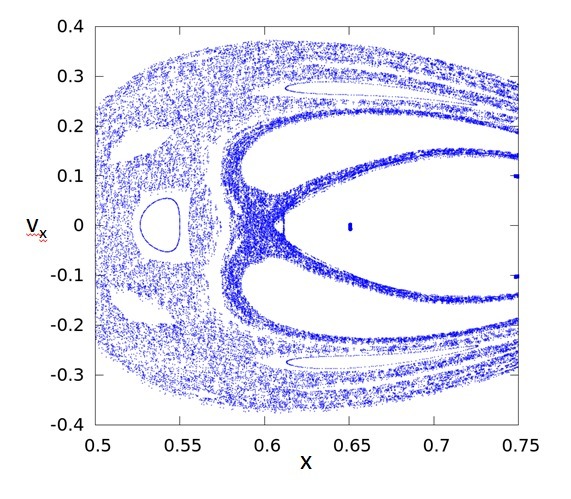}
\caption{\textit{Poincaré map with a zoom.}  Left panel: projection on
  the phase plane $(x,v_x=\dot x)$ of several orbits of the Poincar\'e map on
  $\Mcal_P$.  The time step is $\dt=T/n_s$ with $n_s=1000$.  The Sun,
  Jupiter and the Lagrange point $L_4$ are indicated by a yellow, brown
  and green square respectively. The blue horizontal line
  is the $x$-axis and the vertical one is parallel to the $y$-axis
  and passing trough the $L_4$ Lagrangian point. The orbits are within a region,
  whose boundary is delimited by red lines defined according to
  equation \ref{eq6} by $v_x=\pm\,\sqrt{x^2
    +2(1-\mu)/|x+\mu|+2\mu/|x-1+\mu|-J}$. Right panel: magnification
  of the region delimited by the purple box in the left panel.}
\label{fig:6}
\end{figure}
The dynamic stability can be analyzed by considering the error on a
set of points of $\Mcal_P$ for a fixed value $n$ of iterations of the
Poincar\'e map.
If we consider the reversibility error no intersection of the orbit with 
the manifold $\Mcal_P$ is required, since
we start from an initial point in $\Mcal_P$ and come back to it up to an error due to the
round-off or random perturbation. Since for  a chaotic orbit the error saturates to
$1$ after a few hundreds iterations, the choice $n=100$ is already adequate to
distinguish regions of regular and chaotic behavior, where the error is separated
by many orders of magnitudes. The  REM error is computed in  a grid of points selected
in a rectangular domain of  the $\left( x,\dot x \right)$ plane to which  corresponds a domain in $\Mcal_P$ having
fixed the value $J$ of the Jacobi invariant.
\space
This method is fast and can be compared, in terms of speed, with the fast Lyapounov indicator (FLI, \cite{froshele}), but
its remarkable propriety is that it can be used to estimate the global error due to the round-off.

\begin{figure}[!ht]
\centering
\includegraphics[width=8 cm, height=8 cm]{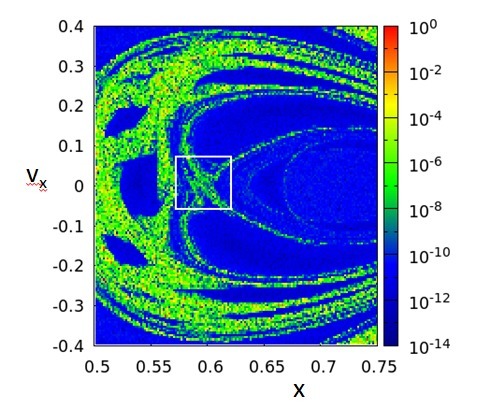}
\includegraphics[width=8 cm, height=8 cm]{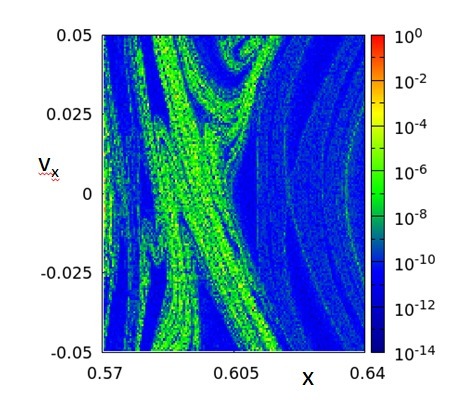}
\caption{\textit{REM color map.}  Left panel: the Reversibility Error
  due to round-off for the one period map $M$. The time step is
  $\dt=T/n_s$ with $n_s=1000$. The number of periods for the computation
  of the REM is $n=1000$.   A color scale is used for the points
  in a regular grid chosen in the same region of the $(x,v_x=\dot x)$ phase
  plane as in the right panel of figure \ref{fig:7}. The presence of regions of
  regular and chaotic motion appears neatly. On the  region of regular motion
  the  error is close to the round-off. On the region of chaotic orbits
  the error is close to $1$.  Right
  panel: the magnification of a transition region corresponding to the white box
  in the left panel.}\label{fig:7}
\end{figure}

In figure \ref{fig:6} we present a portrait of the whole phase space,
 delimiting with red lines the allowed region, defined by equation \ref{eq6},
 and its magnification.  In figure \ref{fig:7} we show the plot, using a color
 scale, of the REM error for the round-off with $n=1000$ iterations of
 the one period map. The chosen phase space region is the same as in figure  
\ref{fig:6} right panel. A new magnification is also shown on the right panel,
 corresponding to the white box on the left panel.
 \spi
 The REM color plot
 allows a rapid and effective visualization of the dynamic
 stability of the system with respect to random perturbations, and
 requires only few lines of code for a given symplectic integrator.
 For the round-off or random perturbations the REM plot requires a
 moderate CPU time even on a fine grid, since the number $n$ of
 iterations is low.

%
%

%
%

\newpage
\section{Fidelity}
\label{sec5}

\spazio The speed at which the dynamic evolution looses memory of the
initial condition is measured by the correlation decay rate. Given an
orbit ${\bm x}_n= M^n({\bm x})$,  where $M$ is a symplectic map and $f(\xbf)$ is an observable
(dynamic variable), one defines the
correlation according to

\begin{equation}\label{eq32}
\setstretch{1.5}
\begin{aligned}
& \hat C(n)  = < f(M^n ) f > - <f>_\mu <f>   \\
& C(n)  = < f(M^n ) f > - <f(M^n)><f>   
\end{aligned}
\end{equation}

Though these  definitions are equivalent in the limit $n \to \infty$, only the second one is suitable for numerical
computations.  The  averages are defined according to 

\begin{equation}\label{eq33}
 <f>= \int_{\cal E} f(\xbf) dm({\bm x})   \qquad \qquad 
 <f>_\mu= \int_{\cal E} f(\xbf) d\mu({\bm x})\equiv
\lim_{n\to \infty }\int_{\cal E}f(M^n({\bm x}) ) \, d m({\bm x})  
\end{equation}
where $m({\bm x})$ denotes the normalized Lebesgue measure. We denote by $m_L$ the Lebesgue measure;
for example $\mu_L(\Bcal)$ is the area of $\Bcal$ if $\Bcal\in \Reali^2$ and the
volume of $\Bcal$ if $\Bcal\in \Reali^4$. If $\Ecal$ is an invariant
domain of phase for any $\Bcal \subset \Ecal$ the measure $m$ is
defined by 
\begin{equation}\label{eq34}
m(\Bcal)={m_L(\Bcal)\over m_L(\Ecal)}
\end{equation}
The invariant measure has the following property 
\begin{equation}\label{eq35}
  \mu(M^{-1}(\Bcal))=\mu(\Bcal).
\end{equation}
If the map is symplectic then its inverse is unique. In this case the invariance condition 
in this case reads $\mu(M^{-1}(\Bcal))=\mu(M(\Bcal))=\mu(\Bcal)$.
For the one period map the phase space is $\Reali^4$. If we
consider an initial set $\Acal_0\subset \Reali^4$ the invariant
manifold $\Ecal$ is the union of all the forward images of
$\Acal_0$. Remark that $\Acal_0$ can be reduced to a single point, in
which case $\Ecal$ is just the orbit having this point as initial
condition.  The Lebesgue measure is just the volume so that
\def\Vol{\,\hbox{Vol}\,}
\begin{equation}\label{eq36}
  \Ecal= \cup_{n=0}^\infty M^n (\Acal_0)
\qquad \qquad  \mu(\Bcal)= {\Vol(\Bcal) \over \Vol(\Ecal)}, \, \qquad \Bcal \subset \Ecal.
\end{equation}
We may compute the fidelity for the one period map, which is the
Poicar\'e map $\tau=0\, \hbox{mod}\, 2\pi$ for the Hamiltonian $H_F$ in
the fixed frame. It is computationally more
convenient to consider the Poincar\'e map $y=0,\,\, p_y>0$ for the
Hamiltonian $H$ in the rotating frame. This is a map define on the 2D manifold
$\Mcal_P$ and its projection on the $\left( x,\dot x \right)$ or $\left( x, p_x \right)$ phase plane has an
invariant measure $\mu$ given by the normalized area with respect to
an invariant domain $\Ecal$. Typically, ${\cal E}$ is the closure of
an orbit issued from a given point or the union of the images of a
given domain, which numerically is sampled with a finite set of
points.  If the points of the orbit $\xbf_0,\xbf_1,\ldots,\xbf_n$ were
random independent variables, the correlation would vanish for any
$n$. For a deterministic Hamiltonian system the correlation does not
decay or decays as $n^{-1}$ for regular orbits whereas it decays
exponentially fast to zero for chaotic orbits.  If the system is
perturbed deterministically, stochastically or by round-off the
perturbed orbit looses memory of the unperturbed one. The Fidelity is
defined as the correlation between the unperturbed orbit and the
perturbed one after $n$ iteration of the perturbed map according to
\begin{equation}\label{eq37}
\setstretch{1.5}
\begin{aligned}
 \hat F_\epsilon(n)  & = < f(M^n ) f(M_{\eps}^n) > - <f>_{\mu} <f>_{\mu_{\eps}}   \\
 F_\epsilon(n) &= < f(M^n )  f(M_{\eps}^n) > - <f(M_\eps^n)><f(M^n)>
\end{aligned}
\end{equation} 
where $\mu_\eps$ is the invariant measure associated to the perturbed
map $M_\eps$, namely the stationary measure in the case of random perturbations. 
  For Hamiltonian systems the invariant and the stationary measures are 
equal to the  normalized  Lebesgue measure  $\mu_\epsilon=\mu=m$ so that 
there is a unique definition  (also for the correlation) and the term to be 
subtracted in equation \ref{eq37} is $\langle \, f\,\rangle^2$.
Another definition of Fidelity  (related to  REM), for symplectic maps, is the following
\begin{equation}\label{eq38}
\setstretch{1.5}
\begin{aligned}
  F^{(R)}_\epsilon(n)  & = <   f(  M_{\eps}^{-n}\circ  M_{\eps}^n)\,\,f > - <f>^2 
\end{aligned}
\end{equation} 

For regular maps such as translations on the torus ${\Bbb T}^d$ the
correlations and the Fidelity do not decay.  For anisochronous maps on
the cylinder ${\Bbb C}={\Bbb T}^d\times {\Bbb I}$ (where ${\Bbb I}$ is
a pluri-interval in $\Reali^d$ to which the actions belong) the
correlations and the Fidelity decay as $1/n$ for observables whose
average on every torus is the same. This behavior is typical of
integrable systems.  For random perturbations depending on $\eps
\xibf$ where $\xibf$ is a vector of independent random variables with
zero mean and unit variance, the Fidelity of $M_\eps $ with respect to
$M$ decays exponentially if $M$ is a regular map
\begin{equation}\label{eq39}
 F_\epsilon(n) \sim  e^{-d_n^2} \sim  \exp \parton {-c\,\epsilon^2 n^{2\beta}}  
\end{equation}
where the exponent $\beta$ is $3/2$ for a generic observable. For an
isochronous system where the angle is stochastically perturbed the
exponent is $\beta=1/2$.  For chaotic maps the Fidelity decays as
\begin{equation}\label{eq40}
F_\epsilon(n) \sim  e^{-d_n^2} \sim  \exp \parton {-c\,\epsilon^2 10^{2n \beta}} 
\end{equation}
If $M$ is the one period map we have $\beta=\lambda\,\,T/\ln 10$
where $\lambda$ is the maximum Lyapunov exponent. If we choose $M$ equal to the
one step map $M_\dt$ then $\beta=\lambda \,\dt/\ln 10$.  If $M$ is
the Poincar\'e map the time between two intersections is comparable
with $T$ so that $\beta\sim \lambda\,\,T \ ln(10)$.  The Fidelity exhibits a
plateau extending from $n=0$ to $n=n_*$ defined by
\begin{equation}\label{eq41}
 n_*={\ln \,\eps^{-1}\over \beta\, \ln 10} 
\end{equation}
followed by a super-exponential decay.  These results where proved for
linear maps on the torus and the cylinder with additive noise
$M_\eps=M+\eps \xi$ \cite{Marie-Chaos}.  Rigorous results in a more
general setting for deterministic perturbations were obtained for
chaotic maps with exponentially decaying correlations \cite{Marie}.
If the perturbation is due to the round-off then the Fidelity does not
decay for regular maps such as the translations on the torus ${\Bbb
  T}^d$.  If the perturbation is a frequency shift linearly depending
on the action, the Fidelity decays as $1/n$.  \spa The Fidelity
behavior when the perturbation is due to the round-off changes
drastically for an integrable system if action angle or Cartesian
coordinates are used. In the first case the forward and reversibility
error do not grow and the Fidelity does not decay. In the second case
the error grows with a power law whose exponent is $\beta=1/2$ if the
map is isochronous, $\beta=3/2$ if it is
anisochronous. Correspondingly the Fidelity has an exponential
decrease with the same exponent $\beta$.  When Cartesian coordinates
are used the map is computationally complex enough that round-off and
random perturbations produce the same effect.  In action angle
variables the round-off is ineffective whereas the random
perturbations cause a power law growth of the global error and an
exponential decay of Fidelity according to equations \ref{eq39} and \ref{eq40}.
We have checked numerically this behavior for harmonic and anharmonic
oscillators.  Non-integrable Hamiltonians exhibit both regular and
chaotic orbits. The Fidelity decay is exponential an super-exponential
respectively and the decay law is the same for the round-off and
random perturbations \cite{Zanlungo_EL}, \cite{Zanlungo} in agreement
with the same exponential growth of the global error described in the
previous section.  \spa For the $3$-body problem we have computed
the Fidelity for the Poincar\'e map in the rotating system because it
is 2D using stochastic perturbations of
amplitude $\epsilon > 10^{-6}$.  In this case linear interpolation can
be used since the error it involves is at least 4 orders of magnitude
below the random error (the use of the H\'enon method is not
straightforward since the integration is carried out in the fixed
reference frame).  

%
\begin{figure}[!ht]
 \centering
 \includegraphics[width=8 cm, height=8 cm]{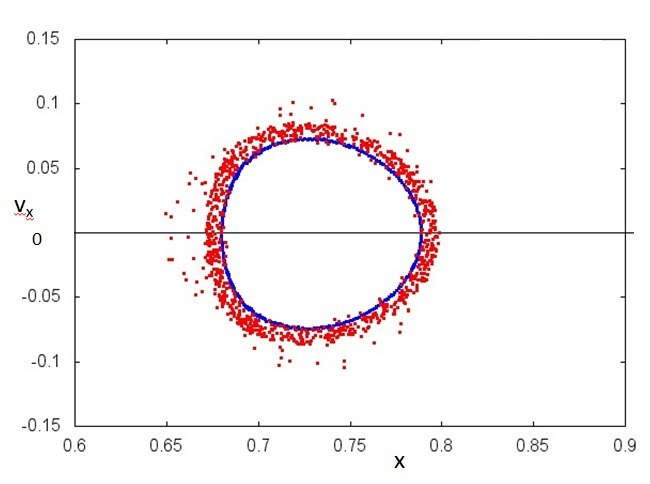}
\includegraphics[width=8 cm, height=8 cm] {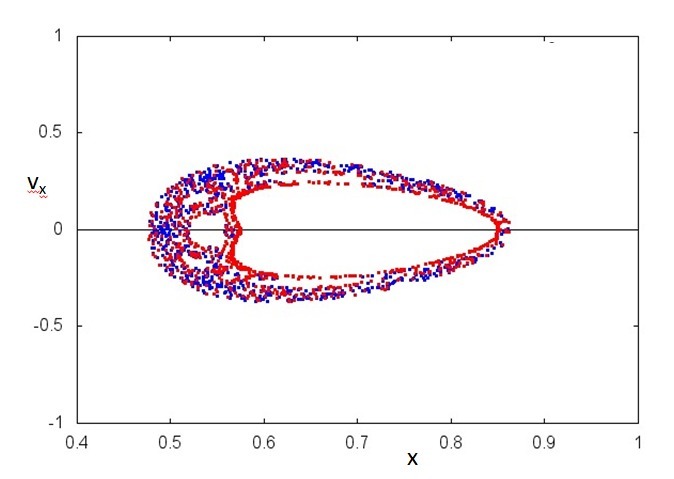}
 \caption{\textit{Unperturbed and noisy orbits.} Left panel:
   projection on the  $(x,v_x=\dot x)$ phase plane of a regular  orbit in the Poincar\'e
   section ($y=0$, $\dot y = 0$) with a stochastic perturbation of amplitude $\eps=10^{-3}$
   (red dots).  The initial point is $x(0)=0.68$ and $\dot x = 0$ in the Jacobi manifold $J=3.07$. 
   The unperturbed orbit is also
   shown (blue dots). Right panel: chaotic orbit with initial points
   $x(0)=0.56$ (blue dots). The  orbit with  a stochastic
   perturbation of amplitude $\eps=10^{-4}$ is shown  (red dots).}
\label{fig:17}
\end{figure}

%

The Jacobi manifold is defined by $J=3.07$ and the
section half plane is $y=0, \,\dot y>0$.  The symplectic perturbation
in this case is introduced in the second order integrator
$M^{(2)}_{\dt}=I+\dt \,N$ by modifying it into $M^{(2)}_{\eps,\dt}=I+\dt
N(1+\epsilon_0 \xibf)$.  By composing three second order maps the
fourth order symplectic map $M^{(4)}_{\eps,\dt}$ is obtained and
the Poincar\'e map $M$ is computed. If $\dt=T/n_s$ then the number of
iterations from two subsequent sections is comparable with $n_s$.
  We have chosen $n_s=100$.   The random vector was changed
only after each section and kept constant until the next section;
changing it at every time step produced no significant difference.

\par\noindent We have analyzed the Fidelity for two distinct initial
conditions on the Poincar\'e section: $x(0)=0.68$ for a non resonant
orbit diffeomorphic to a circle, and $x(0)=0.56$ for chaotic
orbit. The initial point $x(0)=0.55$ considered in the previous
section belongs to a stable resonant orbit formed by three islands.
The orbits in the phase plane $\left( x,\dot x \right)$ with and without the
stochastic perturbation are shown in figure \ref{fig:17}.
\\
For the
regular orbit we consider a sequence of values of the noise amplitude
$\eps_0= 2^{1-m}\,10^{-3}$ for $0\leq m\leq 4$.  The effective
perturbation of the map is $\epsilon_0\,\dt=\epsilon_0\,2\pi/n_s$.
The Fidelity exhibits an exponential decay which can be fitted by
\begin{equation}\label{eq42}
 F_\epsilon(n)= F_\epsilon(0) \exp (-C(\eps) \,n^3).
\end{equation}
%
 \begin{figure}[!htb]
\centering  
 \includegraphics[width=8 cm, height=8 cm]{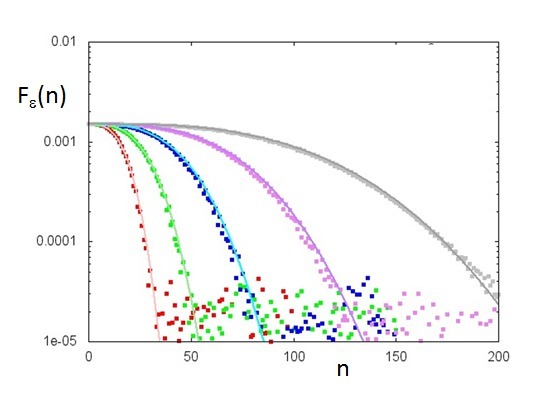} 
 \includegraphics[width=8 cm, height=8 cm]{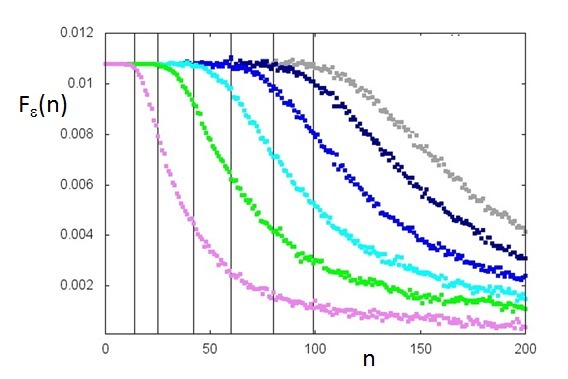}
 \caption{\textit{Fidelity for a regular and chaotic orbit with
     noise.}  Left panel: plot of the Fidelity $F_\eps(n)$ for the
   observable $f=x$ in the case of a regular orbit with initial point
   $x(0)=0.68$ and $\dot x(0)=0$, stochastically perturbed. The small squares
   are the results of simulations with different values of the
   stochastic perturbation amplitude: $\eps= 2\,10^{-3} $ red squares,
   $\eps= 10^{-3} $ green squares, $\eps= 5\,10^{-4} $ blue squares,
   $\eps= 2.5\,10^{-4} $ purple squares, $\eps= 1.25\,10^{-4} $ gray
   squares. The number of steps per period for the symplectic fourth order integrator is
   $n_s=100$. The number of realizations of the stochastic perturbation is $N=100$.  
   The continuous lines are the fits according to equation $exp(-C(\eps)n^{2 \beta})$. The fitted
   coefficients $C(\eps)$ exhibit a quadratic dependence on $\eps$ 
   namely $C(\eps)=c\eps^2$ in agreement with equation \ref{eq39}.
   Right panel: plot of the Fidelity $F_\eps(n)$ for a chaotic orbit
   with initial point $x(0)=0.56$, $\dot x(0)=0$ stochastically
   perturbed.  The perturbation amplitude is: $\eps= 10^{-4} $ purple
   squares, $\eps= 10^{-6} $ green squares, $\eps= 10^{-8} $ cyan
   squares, $\eps= 10^{-10} $ blue squares, $\eps= 5\,10^{-12} $ dark
   blue squares, $\eps= 10^{-14} $ gray squares.  The vertical lines
   correspond to the end of the plateaus $n_*(\epsilon)$, which depend
   linearly on $\ln (1/\epsilon)$.  }
 \label{fig:18}
\end{figure}
In figure \ref{fig:18} left we show the Fidelity for the observable
$f(\xbf)=x$ and the result of a least squares fit according to
equation \ref{eq42}. The coefficients $C(\epsilon)$ obtained from the fit
exhibit a quadratic dependence on $\epsilon$ according to $C(\eps)= c\eps^2 $ 
with $c\simeq 30$.  In a previous paper \cite{Marie-Chaos},
the decay rate $\exp(-\eps^2\,n^3)$ was proved to occur for a
stochastically perturbed map of the cylinder for an observable
$f(\theta)$ where $\theta $ is the angle variable.  In the $3$-body
problem the unperturbed orbit is close to a circle and the radial
diffusion of the perturbed orbit shows that the perturbation affects
also the action variable $\jmath$.  As a consequence since $x\simeq
(2\jmath)^{1/2}\,\cos\theta$ the observed decay law is compatible with
the result proven for the stochastically perturbed map of the
cylinder.
For the chaotic orbit we compute the Fidelity for the sequence of
values of the noise amplitude $\eps_0=10^{-2m}$ for $2\leq m\leq 7$
though only for $m\le 4$ the error in the linear interpolation can be
safely neglected. In figure \ref{fig:18} right we show the plots of the
Fidelity $F_\epsilon(n)$. For any value of $\eps$ the Fidelity
exhibits a plateau up to a value $n_*(\eps)$.  The plateaus followed
by a super-exponential decay are observed in other chaotic maps and
the result was rigorously proved for the Bernoulli maps
\cite{Marie-Chaos}, where $n_*$ was given by equation \ref{eq41}. In the
present case the growth of $n_*$ is linear with $\ln(1/\eps)$ to a very
good approximation.
  
%
%

%
%
\def\spi{\vskip 0.05 truecm \noindent }

\section{ Conclusions}
\label{sec6}
In this paper we have applied the Reversibility Error Method (REM) and the
Fidelity analysis to investigate the dynamic stability of the
restricted three body problem.  The perturbation is due either to the
round-off or to random errors.  The combined use of REM and Fidelity
appears to be adequate to explore the dynamical features on a given
invariant Jacobi manifold.  The reversibility error provides
asymptotically the same information as the forward error but does not
require the exact computation of the unperturbed orbit. Therefore, is well suited to
inspect the effect of round-off.  The loss of memory of the perturbed
orbit it is measured by the Fidelity decay whose computation requires a
Monte-Carlo sampling of the unperturbed orbits.  The round-off appears
to produce the same effects as random perturbations, if a single
realization is considered, provided that the map is sufficiently
complex from the computational viewpoint. In the case of random
perturbations a smoother asymptotic behavior of the error is achieved
by averaging over several realizations.  The computation of the
reversibility error for a fixed number of iterations on a grid of
points on a 2D manifold, combined with a color plot for visualization, is a
straightforward procedure which allows to explore the dynamic
stability of the map, especially in the transition regions. 
To distinguish regular from chaotic orbits very low value of $n$ can be used ($n<100$)
just as for the Fast Lyapounov Indicator (FLI).
The maximum Lyapunov exponents and related indicators require more elaborate
algorithms and extrapolations to infinity. The remarkable property 
of the REM method allow to quantify the global error due to the round-off.  \spi The Fidelity is
computationally expensive but provides statistical information.  For
regular orbits with a random perturbation of amplitude $\epsilon$ the
REM and FEM global errors growth follows a power law $\epsilon
\,n^\beta$ whereas the Fidelity exhibits an exponential decay
$e^{-c\,\epsilon^2\,n^{2\beta}}$.  For chaotic orbits with a random
perturbation the REM and FEM global error growth is $\epsilon \,\,
{10}^{n\,\beta }$, whereas the Fidelity decay is $\exp(-c\,\epsilon^2
\,\,{10}^{2n \beta  })$.	
\spi
For the symplectic map used to
integrate the $3$-body problem the asymptotic behavior of REM and
FEM global errors due to a random perturbation and the corresponding
decay of Fidelity are fully confirmed.  We have observed that the REM
global errors  growth and the Fidelity decay for the round-off are asymptotically the
same as for random perturbation, even though the uncertainty on the exponent
$\beta$ is larger.  This result is expected if the map is
complex enough from  the computational viewpoint.  For an integrable
system the result is different when action angle variables are used,
since the computational complexity is too low.  Indeed with round-off the REM error
vanishes and the Fidelity does not decay,  whereas with a random
perturbation the previous scaling laws are satisfied.
\spi
To summarize we claim that REM and Fidelity appear to be adequate to
analyze the dynamic stability of non-integrable systems with a few
degrees of freedom.  Their use may be recommended to explore the
transition regions where regular and chaotic dynamics coexist and
their relative weight affects the statistical properties, as observed
and proved for simple models, in the case of Poincar\'e recurrences
\cite{Hu}.  The results obtained for the $3$-body problem suggest
the application of REM and Fidelity to the few body problem as the
next natural step.  The few body problem is a key issue in the
description of observed exo-planets (\cite{pulsar}). Indeed the stability of planets
in the habitable zone is a necessary, tough not sufficient, condition for the existence of
extraterrestrial life \cite{HabZone}, \cite{binary}. The presence of MMRs can stabilize or 
destabilize the orbit of a planet and the consequent evolution of the planetary system (see for instance \cite{morbidelli}). The
recent progress in the field of planetary science, due to the Kepler \cite{2014PNAS..11112647B} and GAIA missions \cite{2015ASPC..496..121E}, give the opportunity to test a lot
of planetary systems including resonant or quasi-resonant exo-planets ( see \cite{fabricky} for a review of known resonant or quasi resonant extra-solar systems). Other fields of
astrophysics which might benefit of the proposed approach are the
characterization of regular and chaotic orbits in elliptical galaxies
\cite{binTre}, the motion of binary black holes at the centre of
galaxies \cite{Merritt} and generic Hamiltonian astrophysical systems
(for example \cite{cincotta}, \cite{Maffione} and reference
therein). One of the future and relevant application in astrophysics
is the study of stochastic orbits in axial-symmetric potentials built
with the technique of holomorphic shift \cite{Ciotti}.

\section*{Acknowledgements}
Federico Panichi gratefully acknowledges support from the Polish
National Science Centre MAESTRO grant DEC 2012/06/A/ST9/00276.



%
%
%
%
 \appendix

\section{Error analysis for 2D linear symplectic maps }
\label{appendix-sec1}
We consider the case of a  $2\times 2$ symplectic  matrix $A$ 
distinguishing three possible cases:
\spa
I) If $ |\Tr(A)|>2$ the eigenvalues are real  and  $A=U\Lambda U^{-1}$
where $\Lambda=\hbox{diag}\,(e^{\lambda}, e^{-\lambda}) $.   Choosing $\det U=1$  we introduce the positive
matrix  
\begin{equation}\label{a1}
V= U^T U= \begin{pmatrix}a & b \\ & \cr b & c \end{pmatrix}    \qquad \qquad a,c>0 \qquad ac-b^2=1    
\end{equation}
The traces of $A^k\,(A^k)^T$ and  $V\Lambda^k\,V^{-1}\Lambda^k$ are  equal 
and we obtain the asymptotic behavior of $d_n$ according to
\begin{equation}\label{a2}
 \Tr(A^k\,(A^k)^T)= ac\,e\,^{2k\lambda}+ac\,e^{-2k\lambda} -2b^2
 \qquad \quad \longrightarrow   \qquad \quad  d_n= 
	C\,\,\epsilon\,e^{n\lambda}+O(n\,e^{-\lambda n})    
\end{equation}
where $C=(ac)^{1/2}\,(1-e^{-2\lambda})^{-1/2}$. This asymptotic
approximation to $d_n$ is valid only for $\lambda$ sufficiently
greater than zero. Taking the limit $\lambda \to 0$ in the exact
expression for $d_n$ we have $d_n=\sqrt{2}\,\epsilon \,n^{1/2}$.
\spa
II) If $ |\Tr(A)|<2$ the eigenvalues are complex of unit modulus
$e^{\pm i\omega }$ so that we can write $A=U R(\omega) U^{-1}$ where
$R$ is the rotation matrix and $U$ is a real matrix real.  Still using
the matrix $V=U^T \, U$ introduced above we evaluate the trace of
$A^k\,(A^k)^T$ and the asymptotic behavior of $d_n$ according to
\begin{equation}\label{a3}
\Tr(A^k\,(A^k)^T)= 2\,\cos^2(k\omega)+(a^2+c^2+2b^2) \, \sin^2(k\omega)
 \quad \longrightarrow \quad 
   d_n=  {a+c\over \sqrt{2}}\, \,\epsilon\, n^{1/2} +O(n^{-1/2}) 
\end{equation}

\spa
III) If $Tr A=2$ then either $A=I$ in which case $d_n=
n^{1/2}$ or $A$ has the Jordan form $A=U\Lambda U^{-1}$ where
$\Lambda=\begin{pmatrix}1 & \alpha \\ 0 & 1 \end{pmatrix}$.  Using the
matrix $V$ defined as above the trace of $A^k\,(A^k)^T$ and the
asymptotic behavior of $d_n$ is given by
\begin{equation}\label{a4}
 \hskip -.75 truecm \Tr(A^k\,(A^k)^T)= a^2\,\alpha^2\,k^2+2   \quad \longrightarrow  \quad   
 d_n =\epsilon\,\parqua{ {a^2\,\alpha^2\over 6} (2n^3+3n^2 +n)+  2n }^{1/2}=
{a\over \sqrt{3}}\,\,\alpha \,\,\epsilon\,n^{3/2}   +O(n^{1/2})  
\end{equation}
The last two cases we have examined correspond to the behavior of an
integrable map.  In particular the second case corresponds to an
isochronous map in cartesian coordinates, the third case to an
anisochronous map in action angle coordinates (when $\alpha=0$ we
recover the isochronous case in action angle coordinates). As a
consequence the distance grows as $n^{1/2}$ for an isochronous system
and as $n^{3/2}$ for an anisochronous system.  The expression of
$d_n^{(R)}$ define in equation  \ref{eq27} and using \ref{a2}, \ref{a3} and \ref{a4} is immediately
obtained.  The sums in \ref{eq27} entering the definitions of $(d_n^{(R)})^2$  and $(d_n)^2$ give exactly the same results, since $A$
and $A^{-1}$ have the same eigenvalues. The contribution of
$\Tr[ A^{-n}\bigl ( A^{-n}\big)^T -I]$ does not change the leading
    term in the asymptotic expressions. As a consequence $d^{(R)}_n=
    \sqrt{2}\,\, d_n$ up to the remainder terms both in the expanding
    and the integrable case.  For the round-off the distances $d_n$
    and $d_n^{(R)}$ exhibit large fluctuations as in the case of
    random errors when a single realization is considered. The expressions for $(d_n^{(R)})^2$  and $(d_n)^2$ 
    given by equations \ref{eq25} and \ref{eq26} involve an average which reduced the fluctuations.
    If the map is computationally sufficiently
    complex then the asymptotic behavior of $d_n$ and $d_n^{(R)}$ for
    the round-off is the same and agrees with the one observed and
    theoretically predictable for random errors.  \spa To support the
    previous results on the 2D maps we propose a simple exercise for
    an integrable Hamiltonian $\Hcal=2\pi \,H(p)$ in angle action
    coordinates $(\phi=2\pi x,\, p)$, The scaled system with
    coordinates $(x,p)$ and Hamiltonian $H(p)$ is defined on the
    cylinder $\Toro\times {\Bbb I}$ where $\Toro$ is the interval
    $[0,1]$ with identified endpoints and ${\Bbb I}=[0,a]$. The
    stochastically perturbed Hamiltonian is $H_\epsilon=
    H(p)+\epsilon_0 p\xi_x(t) -\epsilon_0 x \xi_p(t)$ where $\xi_x(t)$
    and $\xi_p(t)$ are independent white noises.
    Letting $\Omega(p)= dH/dp$
     the equations of motion and their solution, up to corrections
    of order $\epsilon_0^2$,  are
\begin{equation}\label{a5}
\begin{aligned}
 \dot x & = \Omega(p)+ \epsilon_0 \xi_x(t)    \hskip 6.5 truecm \dot p=  \epsilon_0 \xi_p(t) \\ \cr
 x & = x_0+\Omega(p_0)\,t + \Omega'(p_0)\, \epsilon_0 \,w_{1\,p}(t)  +\epsilon_0 w_x(t)  \hskip 2 truecm
    p= p_0+\epsilon w_p(t)     
\end{aligned}
\end{equation}
where $w(t)=\int_0^t \,\xi(s) \,ds $ denotes the Wiener noise and
$w_1(t)= \int_0^t \,w(s) \,ds$.  The result, based on the first order
Taylor expansion of $\Omega(p)$, is valid as long as $ \epsilon
\,|w_p(t)| \ll p_0$ namely for $\epsilon t^{1/2}\ll p_0$.  In this case
the distance growth after averaging on the process is
\begin{equation}\label{a6}
 \hskip -.25 truecm   d(t)= \parqua {\Bigl \langle(x-\langle x \rangle )^2 + (p-\langle p \rangle )^2 \Bigr \rangle }^{1/2}=
\epsilon_0\,\,\parqua{(\Omega'(p_0))^2  \,{t^3\over 3} +2  t }^{1/2}  
\end{equation}
The map  $M$ which integrates the previous equation is 
\begin{equation}\label{a7}
 x_n = x_{n-1}+\Omega(p_n)\dt + \epsilon_0 \sqrt{\dt}\,\xi_{x,\,n} \hskip 2 truecm
    p_n= p_{n-1}+\epsilon_0 \sqrt{\dt}\,\,\xi_{p,\,n}  
\end{equation}
where $\xi_{x,\,n}, \,\xi_{p,\,n}$ are independent random variables with zero mean and unit variance. 
Notice that the   amplitude of noise in the symplectic integrator is $\epsilon= \epsilon_0 \,\,(\dt)^{1/2} $.
The map is non linear but  $DM=A$ is constant since $p_n=p_0$ on the unperturbed trajectory.  As a consequence   $DM=A$
has the Jordan  form 
$ A= \left (\begin{array} {cc} 1 &  \alpha   \cr    0 & 1  \cr  \end{array} \right ) $
with $\alpha= \Omega'(p_0)\Delta t$. Letting $\dt \to 0$ keeping
$t=n\dt$ finite the same result as \ref{a6} is obtained.  \spi The
$n^{3/2}$ growth of the forward error due to round-off has been
observed for an integrable system in the specific case of central
motion with $-1/r$ potential \cite{Hairer} and explained by assuming
that the round-off behaves as a random perturbation .
\spi
For a
generic system there is a smooth transition from the $n^{1/2}$ to the
$n^{3/2}$ growth law as the aniso\-chro\-ni\-ci\-ty increases
continuously starting from zero.  A power law approximation
$d_n=c \eps \,n^\gamma$ using the least squares fit to $d_n$ given by 
equation \ref{eq29} (which reduces to equation \ref{a3} in the isochronous case
and to \ref{a4} in the anisochronous case)
with $t=n\dt$, provides an exponent $\gamma(\alpha)$  which
smoothly varies between the asymptotic values $1/2$ to $3/2$ with a
transition at $\alpha\sim 1$.  We have checked this numerically for
the anharmonic oscillator whose Hamiltonian in Cartesian coordinates
is $H=(p^2+x^2)/2+\eta x^4/4$.  Choosing $M$ to be the fourth order
symplectic integrator map with a random perturbation we have computed
with a least squares fit the power law exponent $\beta$ as a function
of $\eta$ for the same initial condition $(x_0,p_0)$.  The dependence
of  $\beta$ on the nonlinearity strength $\eta$ obtained by fitting
the analytic expression $d_n$ given
by equation \ref{eq29} is quite similar.  When the error is due to the round-off
$\beta(\eta)$ has large fluctuations, just as for a single realization
of random errors, but the asymptotic limits and the transition region
are the same.

\section{Second order symplectic integrators}
\label{appendix-sec2}
 The coordinates and velocities transformation from the rotating  to the fixed frame and vice versa 
 read   
 \begin{equation} 
 \begin{pmatrix}
  x_F\\ 
 y_F\\
 \dot{x}_F\\
 \dot{y}_F\end{pmatrix}
 = \begin{pmatrix}
 R(-t) & 0\\ 
 \\
 -\dot{R}(-t) & R(-t) \end{pmatrix}
 \begin{pmatrix}
  x\\ 
 y\\
 \dot{x}\\
 \dot{y}\end{pmatrix} \qquad \qquad \begin{pmatrix}
  x\\ 
 y\\
 \dot{x}\\
 \dot{y}\end{pmatrix}
 = \begin{pmatrix}
 R(t) & 0\\ 
 \\
 \dot{R}(t) & R(t) \end{pmatrix}
 \begin{pmatrix}
 x_F\\ 
 y_F\\
 \dot{x}_F\\
 \dot{y}_F\end{pmatrix}.
 \label{Transf_a}
 \end{equation}
 For brevity we denote the previous transformations as 
 \begin{equation} 
 {\bm{x}}(t)= {\bf{R}}(t){\bm{x}^F}(t)  \qquad \qquad 
  {\bm{x}}^F(t)= {\bf{R}^{-1}}(t) {\bm{x}}(t)
 \label{Transf_b}
 \end{equation}
 
%
%
%
 The second order integrator of Hamilton's equations in the fixed frame
 corresponding to the operator $e^{\dt/2\,D_{V_F}}\, e^{\dt\,D_T{T_F}
 }\, e^{\dt/2\,D_{V_F}}$ explicitly read
 \begin{equation}
 \begin{aligned}
 & x_{F,\,k+1}=x_k+     
 \left[p_{x\,F,\,k}\Delta t + f_x(x_{F,\,k},y_{F,\,k},\tau_k) 
 \frac{(\Delta t)^2}{2}\right]      \\ 
 & y_{F,\, k+1}=y_{F,\,k}+   p_{y\,k}\,\Delta t + f_y(x_{F,\,k},y_{F,\,k},\tau_k)\frac{(\Delta t)^2 }{2}    \\
 & \tau_{k+1}=\tau_k+\Delta t    
 \end{aligned}
 \end{equation}
 followed by
 \begin{equation}
 \begin{aligned}
  p_{x\,F\,k+1} & = p_{x\,F,\,k}+     
 \Bigl[f_x(x_{F,\,k},y_{F,\,k},\tau_k) \; + \;
  f_x(x_{F,\,k+1},y_{F,\,k+1} ,\tau_{k+1}) \Bigr ]   \frac{\Delta t}{2}      \\
  p_{y\,F,\,k+1} & =p_{y\,F,\,k} + \Bigl [f_y(x_{F,\,k},y_{F\,k},\tau_k)+f_y(x_{F,\,k+1},y_{F,\,k+1} ,
 \tau_{k+1})\Bigr ] \frac{\Delta t}{2}    \\  
  p_{\tau\,k+1} & =p_{\tau\,k}+ \Bigl[  f_\tau(x_{F,\,k},y_{F\,k},\tau_k)+f_\tau(x_{F,\,k+1},y_{F,\,k+1} ,
 \tau_{k+1})  \Bigr ] \frac{\Delta t}{2} 
 \end{aligned}
 \end{equation}
 %
 where $f_x,f_y,f_{\tau} $ are the derivative of $-V_F(x_F,y_F,\tau)$, defined in equation \ref{eq1}, with  respect to $x_F,y_F,\tau$.
 The  second order integrator of Hamilton's equations in  the rotating  frame corresponding to the operator $e^{\dt/2\,D_V}\, e^{\dt\,D_T} \, e^{\dt/2\,D_V}$ explicitly read  
 %
 \begin{equation}
 \begin{aligned}
  \begin{pmatrix} x_{k+1} \\ \cr  y_{k+1} \end{pmatrix} & =\Rop(\dt) \begin{pmatrix} x_k+p_{x\,k}\dt +f_x(x_k,y_k) (\dt)^2/2  \\ \cr
  y_k+p_{y\,k}\dt +f_y(x_k,y_k) (\dt)^2/2  \end{pmatrix}   
  \\   \cr
 \begin{pmatrix}  p_{x\,k+1}  \\ \cr  p_{y \, k+1}  \end{pmatrix}  & = \Rop(\dt)  \begin{pmatrix}  p_{x\,k} +f_x(x_k,y_k) \dt/2  \\ \cr
 p_{y\,k} +f_y(x_k,y_k) \dt/2 \end{pmatrix}  + \begin{pmatrix}  f_x(x_{k+1}, y_{k+1} ) \\  \cr  f_y(x_{k+1}, y_{k+1} ) \end{pmatrix} 
 \,{\dt\over 2}
 \end{aligned}
 \end{equation}
 where $f_x=-\partial V/\partial x,\,\,  f_y=-\partial V/\partial y$. 
 

\section{maximum Lyapunov Characteristic Exponent: mLCE}
\label{appendix-sec3}

We consider a symplectic map $M$ and the nearby orbits with initial
points $\xbf_0$ and $\xbf_0+\epsilon \wbf_0$ where $\Vert
\wbf_0\Vert=1$.  The evolution is given by $\xbf_n= M^n(\xbf_0)$ and
$\xbf_n+\epsilon \wbf_n= M^n(\xbf_0+\epsilon \wbf_0)$.  The Lyapunov
exponent is defined by
$$ \lambda= \lim_{n\to \infty }\lim_{\epsilon \to 0 }\,\,{1\over n}  \log \,d_n $$
where $\epsilon d_n$ is the orbit divergence at step $n$ 
$$ d_n= \Vert \wbf_n\Vert=\Vert DT^n(\xbf_0) \wbf_0\Vert +O(\epsilon) $$
Given an invariant ergodic component of the constant energy manifold
the sequence convergences to the mLCE $\lambda$
for all the initial directions $\wbf_0$ of the initial perturbation
except for a set of measure zero corresponding to the eigenvectors of
the smallest Lyapunov exponent.  For chaotic orbits the growth of the
distance is $d_n\propto e^{\lambda\,n }$ and consequently it rapidly
reaches the diameter of the invariant sub-manifold. In order to avoid
this a renormalization procedure has to be used. The procedure is the
following.  Letting $\ybf_n$ and
\def\eps{\epsilon}
\def\yren#1{ \,(\ybf_{#1})_R\,}
$\yren{n}$   be the nearby orbit and the re-normalized nearby  we have starting from $n=1$

\begin{equation}
\setstretch{1.5}
\begin{aligned}
& \xbf_1  = M(\xbf_0) \\
 & \ybf_1\equiv \xbf_1+\eps\,\wbf_1  =  M(\xbf_0+\eps\,\wbf_0) = \\
& \quad =  \xbf_1+ \eps \wbf_1 +O(\epsilon)  \\
& \wbf_1 = \eps DM(\xbf_0) \wbf_0,
\end{aligned}
\end{equation}

and at this step 
\begin{equation}
\setstretch{1.5}
\begin{aligned}
& \yren{1} =  \ybf_1  \\
& (d_1)_R  = d_1 
\end{aligned}
\end{equation}
then at the second step
\begin{equation}
\setstretch{1.5}
\begin{aligned}
& \xbf_2 =  M(\xbf_1) \\
& \ybf_2\equiv  \xbf_2+\eps \wbf_2  = M(\xbf_1+\eps\,\wbf_1)  = \\
& \quad =   \xbf_2+ \eps DM(\xbf_1) \wbf_1 +O(\eps^2)  
\end{aligned}
\end{equation}
and the renormalized vector  $ \yren{2} $ is defined by 
\begin{equation}
\setstretch{1.5}
\begin{aligned}
& \yren{2}= M\parton{\xbf_1+ \eps {\wbf_1\over d_1}}=  \xbf_2+\eps {\wbf_2\over d_1 }+O(\eps^2)  \\ 
& (d_2)_R= {1\over \eps}\Vert\yren{2} -\xbf_2  \Vert = {d_2\over d_1} +O(\eps).
\end{aligned}
\end{equation}
As a consequence $ d_2=(d_2)_R \,(d_1)_R$. In general at step  $n$ we have 
\begin{equation}
\setstretch{1.5}
\begin{aligned}
& \xbf_n= M(\xbf_{n-1}) \\
& \ybf_n\equiv  \xbf_n+\eps \wbf_n=  
M(\xbf_{n-1}+\eps\,\wbf_{n-1})  = \\
& \xbf_n+ \eps DM(\xbf_{n-1}) \wbf_{n-1}  +O(\eps^2),
 \end{aligned}
\end{equation} 
and the renormalized vector  $ \yren{n} $ is defined by 
\begin{equation}
\setstretch{1.5}
\begin{aligned}
 \yren{n} &= M\parton{\xbf_{n-1}+ \eps {(\wbf_{ n-1})_R\over (d_{n-1})_R}  }=  \\
 & =  \xbf n+\eps \,DM(\xbf_{n-1}){\wbf_{n-1}\over (d_{n-1})_R }
+O(\eps^2)  = \\
& = \xbf_n+\eps (\wbf_n)_R+O(\eps^2). 
 \end{aligned}
\end{equation} 
It follows that 
\begin{equation}
\setstretch{1.5}
\begin{aligned}
 (\wbf_n)_R & =  DM(\xbf_{n-1}){\wbf_{n-1}\over (d_{n-1})_R }+O(\eps)= \\ 
&={ DM(\xbf_{n-1} DM(\xbf_{n-2})\cdots DM(\xbf_1) DM(\xbf_0)   \wbf_0
\over (d_{n-1})_R  (d_{n-2})_R \cdots (d_1)_R  }+O(\eps) =\\
& = {\wbf_n \over( d_{n-1})_R  (d_{n-2})_R \cdots (d_1)R   } +O(\eps).
\end{aligned}
\end{equation}

The final result reads
$$ d_n=\Vert \wbf_n\Vert=   (d_{n-1})_R  (d_{n-2})_R \cdots (d_1)_R +O(\eps), $$
and the mLCE is expressed by
$$ \lambda =  \lim_{n\to \infty }\,\,\lim_{\eps \to 0 } \, 
\,{1\over n}\,\sum_{j=1}^{n-1} \,\log (d_j)_R.   $$
The algorithm is extremely simple and is expressed by the recurrence 
\begin{equation}
\setstretch{1.5}
\begin{aligned}
& \eps(\wbf_{n-1})_R=  (\ybf_{n-1})_R   - \xbf_{n-1} \\
& (d_{n-1})_R= \Vert \wbf_{n-1}  \Vert \\
&  (\ybf_n)_R = M\parton{\xbf_{n-1} +\eps { (\wbf_{n-1}\over (d_{n-1})_R}}
\end{aligned}
\end{equation} 
initialized by
\begin{equation}
\setstretch{1.5}
\begin{aligned}
& \eps (\wbf_1)_R = \eps \wbf_1 = M(\xbf_0 +\eps \wbf_0) -\xbf_1   \\
& (\ybf_1)_R= M(\xbf_0+\eps \wbf_0) \\ 
& (d_1)_R = d_1=\Vert \wbf_1  \Vert.
\end{aligned}
\end{equation}



\begin{thebibliography}{9}



	
\bibitem[Aarseth (2003)]{AarsethBook}  S. J. Aarseth [2003] {\it Gravitational N-body Simulations: Tools and Algorithms},
Cambridge Monographs on Mathematical Physics.   
\bibitem[Aarseth et al. (Eds.) (2008)]{NbodyLecture}  S. J. Aarseth \& C. A. Tout \& R. S. Mardling [2003] {\it The Cambridge N-Body Lectures, Lect. Notes Phys. 760},  
Springer, Berlin Heidelberg 2008.
\bibitem[Hiroshi et al. (1991)]{SymplNbody}	K. Hiroshi \& H. Yoshida \& N. Hiroshi [1991] {\it Symplectic integrators and their application to dynamical astronomy},   
Celestial Mechanics and Dynamical Astronomy vol. 50, pp. 59-71.
\bibitem[Dehnen \& Read (2011)]{SymplNbody2}  W. Dehnen \& J. I. Read  [2011] {\it N-body simulations of gravitational dynamics},
The European Physical Journal Plus vol. 126.   
\bibitem[Zwart \& Boekholt (2014)]{errors}  S. P. Zwart \& T. Boekholt  [2014] {\it On the minimal accuracy required for simulating self-gravitating systems by means of direct N-body methods},
The Astrophysical Journal Letters vol. 785.     
\bibitem[Dejonghe \& Hut (1986)]{olderror}   H. Dejonghe \& P. Hut   [1986] {\it Round-off sensitivity in the N-body problem},
The Use of Supercomputers in Stellar Dynamics, Lecture Notes in Physics Volume 267, pp. 212-218.   
\bibitem [Hairer (2002)] {Hairer} [2002], E. Hairer, C. Lubich \& G. Wanner  {\it Geometric Numerical Integration. Structure-Preserving Algorithms for Ordinary Differential Equations}, Springer Series in Computa-
tional Mathematics Vol. 31 (Springer, New York, 2002).
\bibitem[Oseledec (1968)] {Oseledec} [1968] Oseledec V. I. [1968] {\it Multiplicative Ergodic Theorem.
 The Lyapunov characteristic numbers of dynamical systems}  (in Russisan). 
Trudy Mosk. Mat. Obsch. vol. 19, pp. 179.210, 1968. Eng1ish 
tras1ation in Trans. Mosc. Math. Soc. vol. 19, pp. 197.    
\bibitem[Benettin (1980)] {Benettin} [1980] G. Benettin G, L. Galgani, A. Giorgilli\&  J. M. Strelcyn  {\it Lyapunov characteristic exponents for smooth dynamical systems and for Hamiltonian systems; A method for computing all
of them. Part 1: theory},  Meccanica, pp. 9-20.  
\bibitem[Benettin (1980)] {Benettin2} [1980]G. Benettin G, L. Galgani, A. Giorgilli \&  J. M. Strelcyn
{\it Lyapunov characteristic exponents for smooth dynamical systems and for Hamiltonian systems; A method for computing all
of them. Part 2: Numerical application}, Meccanica, pp. 21-30.  
\spa
\bibitem[Skokos (2010)] {Skokos} [2010] C. Skokos  {\it The Lyapunov Characteristic Exponents and Their Computation}, 
Lecture Notes in Physics, Berlin Springer Verlag.   
\bibitem[Arnold (2002)] {Arnold} [2002]  V. I Arnold, A I. Neishtadt \&  V. Kozlov  {\it Mathematical aspects of classical
and celestial mechanics } {Encyclopedia of Mathematical Sciences} Vol III Dynamical systems.
 Springer (Original Russian edition  published by  URSS, Moscow 2002).   
\bibitem[Aarseth \& Zare (1974)]{aaresath74}  S. J. Aarseth \& K. Zare  [1974] {\it A regularization of the three-body problem},
Celestial Mechanics, vol. 10, p. 185-205.
\bibitem[Minesaki (2013)]{reversTest}   Y. Minesaki   [2013] {\it Accurate orbital integration of the general three-body problem based on the D'Alambert-type scheme},The Astronomical Journal vol. 145, Issue 3, pp. 14. 

\bibitem[Hollander \& De Luca (2004)]{chaosJ} E. B. Hollander \& J. De Luca [2004]{\it  Regularization of the collision in the electromagnetic two-body problem},Chaos: An Interdisciplinary Journal of Nonlinear Science vol. 14, pp. 1093.


\bibitem[Bakker et al. (2011)]{analRev}   L. F. Bakker \& T. Ouyang \& D. Yan \& S. Simmons   [2011] {\it Existence and stability of symmetric periodic simultaneous binary collision orbits in the planar pairwise symmetric four-body problem},
Celestial Mechanics and Dynamical Astronomy vol. 110, Issue 3, pp. 271-290. 

\bibitem[Casati (1986)] {Casati}  G. Casati, B. V. Chirikov, I. Guarneri \&  D. L. Shepelyansky, {\it Dynamical Sta-
bility of Quantum ”Chaotic” Motion in a Hydrogen Atom}, Phys. Rev. Lett., vol. 56(23),  pp. 2437-
2440.

\bibitem[Marie (2007)]{Marie}	C. Liverani, P. Marie, S. Vaienti [2007]
{\it Random classical Fidelity }  Journal of Statistical Physics, vol.  128, pp.  1079 (2007).
\bibitem[Marie {\it et al.}(2009)]{Marie-Chaos} P. Marie,
  G. Turchetti, S. Vaienti \& F. Zanlungo [2009] {\it Error distribution in randomly perturbed orbits.}, Chaos, vol. 19, 2009.
\bibitem[Faranda (2012)]{Mestre} D. Faranda,
  F.M. Mestre \& G. Turchetti [2012] {\it Analysis of round-off errors
    with Reversibility test as a dynamical indicator}, International
  Journal of Bifurcation and Chaos, vol. 22, Issue 09, 2012.
\bibitem[Turchetti (2010)]{Zanlungo_EL} G. Turchetti,
  S. Vaienti \& F. Zanlungo [2010] {\it Relaxation to the asymptotic distribution of
    global errors in the numerical computations of dynamical systems},
  Europhysics Letters, vol. 89, 40006-40010.
\bibitem[Turchetti (2010)]{Zanlungo} G. Turchetti,
  S. Vaienti \& F. Zanlungo [2010] {\it Relaxation to the asymptotic distribution of
    global errors in the numerical computations of dynamical systems},
  Physica A: Statistical Mechanics and its Applications, vol. 389,
  pp. 4994-5006.
\bibitem[Mauri \& Hannu (2005)]{Valtonen} M. Valtonen \& K. Hannu [2005] {\it The Three-Body Problem},
Cambridge University Press.
\bibitem[Szebehely (1972)]{Sz72} V. Szebehely [1972] {\it The General and Restricted
problems of three bodies}, Springer-Verlag Wien, New York.
\bibitem[Turchetti (1972)]{turchetti} G. Turchetti [1999] {\it Dinamica Classica dei sistemi Fisici}, Ed. Zanichelli, Bologna (out of press), chap. 23, pp. 441 allowable from  \url{http://www.physycom.unibo.it/libro.php}
\bibitem[Murray \& Dermott (1999)]{MD99} C.D. Murray \& S.F. Dermott
  [1999] {\it Solar System Dynamics}, Cambridge University Press.
\bibitem[Yoshida (1990)]{Yoshida} H. Yoshida [1990] {\it Construction of higher order symplectic integrators},
PHYSICS LETTERS A  vol. 150, pp. 262-268. 
\bibitem[Henon (1982)] {Henon} M. H\'enon [1982]   {\it On the numerical computation of Poincar\'e maps}, 
Physica vol. 5D, pp. 412-414.
\bibitem[Froeschlé (1997)]{froshele} Cl. Froeschlé, R. Gonczi1 \& E. Lega1 [1997]{\it The fast Lyapunov indicator: a simple tool to detect weak chaos. Application to the structure of the main asteroidal belt}  Planetary and Space Science, vol. 45, Issue 7, pp. 881-886.
\bibitem[Hu {\it et al.}(2004)]{Hu} H. Hu, A. Rampioni, L. Rossi, G. Turchetti \& S. Vaienti  [2004] {\it Statistics of Poincaré recurrences for area preserving maps with integrable and ergodic components}, Chaos, vol. 14, pp. 160-171.   

\bibitem[Wolszczan \& Frail (1992)]{pulsar} A. Wolszczan \& D. A. Frail [1992] {\it A planetary system around the millisecond pulsar PSR1257 + 12},Nature vol. 355, pp. 145-147. 
\bibitem[Kopparapu et al. (2014)]{HabZone} R. K. Kopparapu \& R. M. Ramirez \& J. SchottelKotte \& J. F. Kasting \& S. Domagal-Goldman \& V. Eymet [2014] {\it Habitable zones around main-sequence stars: dependence on planetary mass},
Nature vol. 355, pp. 145-147. 
\bibitem[Quintana et al. (2007)]{binary} E. V. Quintana \& F. C. Adams \& J. J. Lissauer \& J. E. Chambers [2007] {\it Terrestrial Planet Formation around Individual Stars within Binary Star Systems},
The Astrophysical Journal vol. 660, Issue 1, pp. 807-822.


\bibitem[Batygin et al.(2015)]{morbidelli} Batygin, K., 
Morbidelli, A., \& Holman, M.~J.\ 2015, Apj, 799, 120 


\bibitem[Batalha (2014)]{2014PNAS..11112647B} Batalha, N.M.\ 2014, 
Proceedings of the National Academy of Science, 111, 12647 
\bibitem[Eyer et al. (2015)]{2015ASPC..496..121E} Eyer, L., Rimoldini, L., 
Holl, B., et al.\ 2015, Astronomical Society of the Pacific Conference 
Series, 496, 121 

\bibitem[Fabrycky et al.(2014)]{fabricky} Fabrycky, D.C., 
Lissauer, J.J., Ragozzine, D., et al.\ 2014, Apj, 790, 146 




\bibitem[Binney \& Tremaine {\it et al.}(1987)]{binTre} J. Binney \& S. Tremaine [1987] {\it Galactic dynamics}, Princeton, NJ, Princeton University Press, 1987, 747 p.
\bibitem[Merritt (2013)]{Merritt}  D. Merritt [2013]{\it Dynamics and Evolution of Galactic Nuclei},
Princeton University Press, 2
\bibitem[Cincotta \& Sim{\'o} (2000)]{cincotta} P. Cincotta \&  C. Sim{'\o} [2000] {\it Simple tools to study global dynamics in non-axisymmetric galactic potentials - I},
  Astron. Astrophys. Suppl. Ser. vol. 147, pp. 205-228.
\bibitem[Maffione (2012)]{Maffione} N.P. Maffione, L. A. Darriba,
  P. M. Cincotta \& C. M. Giordano [2012]{\it Chaos detection tools:
    application to a self-consistent triaxial model}, arXiv:1212.3175.
\bibitem[Ciotti (2007)]{Ciotti} L. Ciotti, G. Gainpieri [2007]{\it Exact density potential pairs
from the holomorphic Coulomb field}   Mon. Not. R. Astron. Soc., vol. 376, pp. 1162-1168. 
 
\end{thebibliography}
\end{document}